\batchmode
\makeatletter
\def\input@path{{"C:/Trabajo laptop/Mis articulos/Finished/PMBM arbitrary clutter/Accepted/"}}
\makeatother
\documentclass[english]{IEEEtran}
\usepackage[T1]{fontenc}
\usepackage[latin9]{inputenc}
\usepackage{array}
\usepackage{multirow}
\usepackage{amsmath}
\usepackage{amsthm}
\usepackage{amssymb}
\usepackage{graphicx}
\PassOptionsToPackage{normalem}{ulem}
\usepackage{ulem}

\makeatletter

\providecommand{\tabularnewline}{\\}

\theoremstyle{plain}
\newtheorem{thm}{\protect\theoremname}
\theoremstyle{definition}
\newtheorem{example}[thm]{\protect\examplename}
\theoremstyle{plain}
\newtheorem{lem}[thm]{\protect\lemmaname}

\usepackage{cite}
\allowdisplaybreaks 

\usepackage[margin=8pt,font=footnotesize]{caption}
\usepackage{algorithm}
\usepackage{algpseudocode}
\usepackage{amsmath}  

\makeatother

\usepackage{babel}
\providecommand{\examplename}{Example}
\providecommand{\lemmaname}{Lemma}
\providecommand{\theoremname}{Theorem}

\begin{document}
\title{Poisson multi-Bernoulli mixture filter with general target-generated
measurements and arbitrary clutter}
\author{Ángel F. García-Fernández, Yuxuan Xia, Lennart Svensson \thanks{A. F. García-Fernández is with the Department of Electrical Engineering and Electronics, University of Liverpool, Liverpool L69 3GJ, United Kingdom (angel.garcia-fernandez@liverpool.ac.uk). He is also with  the ARIES Research Centre, Universidad Antonio de Nebrija,  Madrid, Spain. Y. Xia and L. Svensson are with the Department of Electrical Engineering, Chalmers University of Technology, SE-412 96 Gothenburg, Sweden (firstname.lastname@chalmers.se).} }
\maketitle
\begin{abstract}
This paper shows that the Poisson multi-Bernoulli mixture (PMBM) density
is a multi-target conjugate prior for general target-generated measurement
distributions and arbitrary clutter distributions. That is, for this
multi-target measurement model and the standard multi-target dynamic
model with Poisson birth model, the predicted and filtering densities
are PMBMs. We derive the corresponding PMBM filtering recursion. Based
on this result, we implement a PMBM filter for point-target measurement
models and negative binomial clutter density in which data association
hypotheses with high weights are chosen via Gibbs sampling. We also
implement an extended target PMBM filter with clutter that is the
union of Poisson-distributed clutter and a finite number of independent
clutter sources. Simulation results show the benefits of the proposed
filters to deal with non-standard clutter. 
\end{abstract}


\section{Introduction}

Multi-target filtering consists of estimating the current states of
an unknown and variable number of targets based on noisy sensor measurements
up to the current time step. It is a key component of numerous applications,
for example, defense \cite{Blackman_book99}, automotive systems \cite{Held22}
and air traffic control \cite{Buelta21}. Multi-target filtering is
usually addressed using probabilistic modelling, with the main approaches
being multiple hypothesis tracking \cite{Reid79}, joint probabilistic
data association \cite{Bar-Shalom75} and random finite sets \cite{Mahler_book14}. 

In detection-based multi-target filtering, sensors collect scans of
data that may contain target-generated measurements as well as clutter,
which refers to undesired detections. For example, in radar, clutter
can be caused by reflections from the environment, such as terrain,
sea and rain, or other undesired objects \cite{Greco14,Skolnik_book01}.
Multi-target filters applied to real data must then account for these
clutter measurements for suitable performance, for instance, sea clutter
and land clutter in radar data \cite{Helgesen22}, false detections
in image data \cite{Cho14,Scheidegger18}, false detections in audio
visual data \cite{Zhao22}, and underwater clutter in active sonar
data \cite{Mellema20,Grimmett21}.

In the standard detection model, clutter is modelled as a Poisson
point process (PPP) \cite{Mahler_book14}. The PPP is motivated by
the fact that, for sufficiently high sensor resolution and independent
clutter reflections in each sensor cell, the detection process can
be accurately approximated as a PPP \cite{Erdinc09}. The PPP is also
convenient mathematically and it can be characterised by its intensity
function on the single-measurement space. 

For the point-target measurement model, PPP clutter and the standard
multi-target dynamic model with PPP birth, the posterior is a Poisson
multi-Bernoulli mixture (PMBM) \cite{Williams15b,Angel18_b}. The
PMBM consists of the union of a PPP, representing undetected targets,
and an independent multi-Bernoulli mixture (MBM) representing targets
that have been detected at some point and their data association hypotheses.
The posterior density is also a PMBM for the extended target model
\cite{Granstrom20} and for a general target-generated measurement
model \cite{Angel21}, both with PPP clutter and the standard dynamic
model. Extended target modelling is required when, due to the sensor
resolution and the target extent, a target may generate more than
one detection at each time step \cite{Granstrom17}. This makes the
data association problem considerably more challenging than for point
targets. The general target generated-measurement model includes the
point and extended target cases as particular cases, and can for example
be used when there can be simultaneous point and extended targets
in a scenario \cite{Angel21}.

If the birth model is multi-Bernoulli instead of PPP, the posterior
density in the above cases is an MBM, which is obtained by setting
the Poisson intensity of the PMBM filter to zero and by adding the
Bernoulli components of the birth process in the prediction step.
The MBM filter can also be written in terms of Bernoulli components
with deterministic target existence,  which results in the MBM$_{01}$
filter \cite[Sec. IV]{Angel18_b}. It is also possible to add unique
labels to the target states in the MBM and MBM$_{01}$ filters. The
(labelled) MBM$_{01}$ filtering recursions are analogous to the $\delta$-generalised
labelled multi-Bernoulli ($\delta$-GLMB) filtering recursions \cite{Vo17,Beard16}. 

All the above recursions to compute the posterior assume PPP clutter,
but other clutter distributions are also important in various contexts,
for instance, to model bursts of radar sea clutter \cite{Watts08}.
For general target-generated measurements and arbitrary clutter density,
the Bernoulli filter was proposed in \cite{Shen18}, and the probabilistic
hypothesis density (PHD) filter, which provides a PPP approximation
to the posterior, was derived in \cite{Clark12}. The corresponding
PHD filter update depends on the set derivative of the logarithm of
the probability generating functional (PGFL) of the clutter process.
A version of this PHD filter update, written in terms of densities,
which is more suitable for implementation than \cite{Clark12}, is
provided in \cite{Shen21}. There are also other PHD filter variants
with non-PPP clutter for point targets, for example, a PHD filter
with negative binomial clutter \cite{Schlangen16}, a second order
PHD filter with Panjer clutter \cite{Schlangen18}, and a linear-complexity
cumulant-based filter for clutter described by its intensity and second-order
cumulant \cite{Clark18,Campbell21}. A cardinalised PHD filter with
general target-generated measurements and arbitrary clutter is derived
in \cite{Shen22}. Another approach is to estimate the states of Bernoulli
clutter generators by including them in the multi-target state \cite{Mahler_book14,Mahler18}.

This paper shows that for general target-generated measurements and
arbitrary clutter density, the posterior is also a PMBM and we derive
the filtering recursion. This contribution extends the family of closed-form
recursions to calculate the posterior to a more general detection-based
measurement model. As a direct result, this paper also derives the
corresponding (labelled or not) MBM and MBM$_{01}$ filters, including
$\delta$-GLMB filters. We also directly obtain the Poisson multi-Bernoulli
(PMB) filter, which can be derived by Kullback-Leibler divergence
minimisation on a target space augmented with auxiliary variables
\cite{Williams15b,Angel20_e}. We also propose a Gibbs sampling algorithm
for selecting global hypotheses with high weights \cite{Hue02,Hue02b,Vo17,Granstrom18b}
for the PMBM filter for point targets and clutter that is an independent
and an identically distributed (IID) cluster process \cite{Mahler_book14}
with arbitrary cardinality distribution. We show via simulation results
the advantages of the proposed filter in two scenarios: one scenario
for point targets and IID clutter with negative binomial cardinality
distribution, and another scenario for extended targets where there
are a finite number of clutter sources.

The rest of this paper is organised as follows. Section \ref{sec:Models-and-overview}
introduces the models and an overview of the PMBM posterior. The PMBM
filter update is presented in Section \ref{sec:PMBM-filter-update}.
The Gibbs sampling data association algorithm for point-target PMBM
filtering with arbitrary clutter is proposed in Section \ref{sec:Data-associations-Gibbs}.
Finally, simulation results and conclusions are provided in Sections
\ref{sec:Simulation-experiments} and \ref{sec:Conclusions}.

\section{Models and overview of the PMBM posterior\label{sec:Models-and-overview}}

This section presents the dynamic and measurement models in Section
\ref{subsec:Models}, and an overview of the PMBM posterior in Section
\ref{subsec:PMBM-posterior}, with its global hypotheses in Section
\ref{subsec:Set-of-global-hypotheses}. 

\subsection{Models\label{subsec:Models}}

A target state is denoted by $x$ and it contains the variables that
describe its current dynamics, such as position and velocity, and
maybe other attributes, such as orientation and extent. The state
$x\in\mathcal{X}$, where $\mathcal{X}$ is a locally compact, Hausdorff
and second-countable (LCHS) space \cite{Mahler_book14}. The set of
targets at time $k$ is $X_{k}\in\mathcal{F}\left(\mathcal{X}\right)$,
where $\mathcal{F}\left(\mathcal{X}\right)$ represents the set of
finite subsets of $\mathcal{X}$. 

Targets move according to the standard multi-target dynamic model
\cite{Mahler_book14}. Given $X_{k}$, each target $x\in X_{k}$ survives
with probability $p^{S}\left(x\right)$ and moves with a transition
density $g\left(\cdot\left|x\right.\right)$, or disappears with probability
$1-p^{S}\left(x\right)$. New targets are born at time step $k$ according
to an independent Poisson point process (PPP) with intensity $\lambda_{k}^{B}\left(\cdot\right)$. 

A measurement state $z\in\mathcal{Z}$ contains a sensor measurement.
The set $Z_{k}\in\mathcal{F}\left(\mathcal{Z}\right)$ of measurements
at time $k$ is the union of target-generated measurements and independent
clutter measurements, modelled by
\begin{itemize}
\item Each target $x\in X_{k}$ generates a set of measurements with density
$f\left(\cdot|x\right)$. 
\item The set of clutter measurements at each time step has density $c\left(\cdot\right)$.
\item Given $X_{k}$, the measurements generated by different targets are
independent of each other and of the clutter measurements. 
\end{itemize}
It should be noted that $f\left(\cdot|x\right)$ is a general density
for target-generated measurements and we can accommodate any probabilistic
model for target-generated measurements. The probability that at least
one measurement is generated from the target (effective probability
of detection \cite{Granstrom12}) is $1-f\left(\emptyset|x\right)$.
For example, the standard point target model in which a target $x$
is detected with probability $p^{D}\left(x\right)$ and generates
a measurement with density $l(\cdot|x)$ \cite{Mahler_book14}, is
obtained by
\begin{align}
f\left(Z|x\right) & =\begin{cases}
1-p^{D}\left(x\right) & Z=\emptyset\\
p^{D}\left(x\right)l(z|x) & Z=\left\{ z\right\} \\
0 & \left|Z\right|>1.
\end{cases}\label{eq:standard_point_measurement}
\end{align}
In the standard extended target model, we can receive more than one
measurement from a target. Specifically, a target $x$ is detected
with probability $p^{D}\left(x\right)$ and, if detected, it generates
a PPP measurement with intensity $\gamma\left(x\right)l(\cdot|x)$,
where $l(\cdot|x)$ is a single-measurement density and $\gamma\left(x\right)$
is the expected number of measurements \cite{Granstrom17,Gilholm05}.
It is obtained by

\begin{align}
f\left(Z|x\right) & =\begin{cases}
1-p^{D}\left(x\right)+p^{D}\left(x\right)e^{-\gamma\left(x\right)} & Z=\emptyset\\
p^{D}\left(x\right)\gamma^{\left|Z\right|}\left(x\right)e^{-\gamma\left(x\right)}\prod_{z\in Z}l(z|x) & \left|Z\right|>0.
\end{cases}\label{eq:standard_extended_measurement}
\end{align}

We can also combine both (\ref{eq:standard_point_measurement}) and
(\ref{eq:standard_extended_measurement}) to model coexisting point
and extended targets, or use any of the probabilistic extended target
models in \cite{Granstrom17}. In addition, the choice of $f\left(\cdot|x\right)$
can also take into account reflectivity models and the propagation
conditions in the environment \cite{Skolnik_book01,Porter19,Ge20b,Narykov22}.

\subsection{PMBM posterior\label{subsec:PMBM-posterior}}

We will show in this paper that, for the dynamic and measurement models
in Section \ref{subsec:Models}, the predicted and posterior densities
are PMBMs. That is, given the sequence of measurements $\left(Z_{1},...,Z_{k'}\right)$,
the density  $f_{k|k'}\left(\cdot\right)$ of $X_{k}$ with $k'\in\left\{ k-1,k\right\} $
is
\begin{align}
f_{k|k'}\left(X_{k}\right) & =\sum_{Y\uplus W=X_{k}}f_{k|k'}^{\mathrm{p}}\left(Y\right)f_{k|k'}^{\mathrm{mbm}}\left(W\right),\label{eq:PMBM}\\
f_{k|k'}^{\mathrm{p}}\left(X_{k}\right) & =e^{-\int\lambda_{k|k'}\left(x\right)dx}\prod_{x\in X_{k}}\lambda_{k|k'}\left(x\right),\\
f_{k|k'}^{\mathrm{mbm}}\left(X_{k}\right) & =\sum_{a\in\mathcal{A}_{k|k'}}w_{k|k'}^{a}\sum_{\uplus_{l=1}^{n_{k|k'}}X^{l}=X_{k}}\prod_{i=1}^{n_{k|k'}}f_{k|k'}^{i,a^{i}}\left(X^{i}\right),\label{eq:MBM}
\end{align}
where $\lambda_{k|k'}\left(\cdot\right)$ is the intensity of the
PPP $f_{k|k'}^{\mathrm{p}}\left(\cdot\right)$, representing undetected
targets, and $f_{k|k'}^{\mathrm{mbm}}\left(\cdot\right)$ is a multi-Bernoulli
mixture representing  targets that have been detected at some point
up to time step $k'$. The sum in (\ref{eq:PMBM}) is the convolution
sum, which implies that the PPP and MBM are independent \cite{Mahler_book14}.
The symbol $\uplus$ denotes the disjoint union and the sum is taken
over all mutually disjoint (and possibly empty) sets $Y$ and $W$
whose union is $X_{k}$.

The MBM in (\ref{eq:MBM}) has $n_{k|k'}$ Bernoulli components (potential
targets), each with $h_{k|k'}^{i}$ local hypotheses. There is a local
hypothesis $a^{i}\in\left\{ 1,...,h_{k|k'}^{i}\right\} $ for each
Bernoulli $i>0$. To handle arbitrary clutter, we also introduce a
local hypothesis $a^{0}\in\left\{ 1,...,h_{k|k'}^{0}\right\} $ for
clutter. Each $a^{i}$ is an index that associates the $i$-th Bernoulli,
for $i\geq0$, or the clutter, for $i=0$, to a specific sequence
of subsets of the measurement set, see Section \ref{subsec:Set-of-global-hypotheses}.
A global hypothesis is denoted by $a=\left(a^{0},a^{1},...,a^{n_{k|k'}}\right)\in\mathcal{A}_{k|k'}$,
where $\mathcal{A}_{k|k'}$ is the set of global hypotheses, see Section
\ref{subsec:Set-of-global-hypotheses}. The weight of global hypothesis
$a$ is $w_{k|k'}^{a}$ and meets
\begin{equation}
w_{k|k'}^{a}\propto\prod_{i=0}^{n_{k|k'}}w_{k|k'}^{i,a^{i}}\label{eq:global_weights}
\end{equation}
where $w_{k|k'}^{i,a^{i}}$ is the weight of the $i$-th Bernoulli
component, or clutter if $i=0$, with local hypothesis $a^{i}$, and
$\sum_{a\in\mathcal{A}_{k|k'}}w_{k|k'}^{a}=1$. A difference with
PMBM filters with PPP clutter \cite{Williams15b,Granstrom20,Angel21}
is that global hypotheses and weights explicitly consider clutter,
with index $i=0$.

The $i$-th Bernoulli component with local hypothesis $a^{i}$ has
a density
\begin{align}
f_{k|k'}^{i,a^{i}}\left(X\right) & =\begin{cases}
1-r_{k|k'}^{i,a^{i}} & X=\emptyset\\
r_{k|k'}^{i,a^{i}}p_{k|k'}^{i,a^{i}}\left(x\right) & X=\left\{ x\right\} \\
0 & \mathrm{otherwise}
\end{cases}\label{eq:Bernoulli_density_filter}
\end{align}
where $r_{k|k'}^{i,a^{i}}$ is the probability of existence and $p_{k|k'}^{i,a^{i}}\left(\cdot\right)$
is the single-target density. It should be noted that in this paper
we use the following nomenclature for Bernoulli components, densities
and local hypotheses: Each Bernoulli component, which is indexed by
$i$ and is initiated by a non-empty subset of measurements at a given
time step (see Section \ref{sec:PMBM-filter-update}), has $h_{k|k'}^{i}$
local hypotheses, each with an associated Bernoulli density, indexed
by $i,a^{i}$. Then, the total number of Bernoulli densities in (\ref{eq:MBM})
across all global hypotheses is $\sum_{i=1}^{n_{k|k'}}h_{k|k'}^{i}$,
and the number of multi-Bernoulli densities is $|\mathcal{A}_{k|k'}|$,
which denotes the cardinality of $\mathcal{A}_{k|k'}$. 

\subsection{Set of global hypotheses\label{subsec:Set-of-global-hypotheses}}

We proceed to describe the set $\mathcal{A}_{k|k'}$ of global hypotheses.
We denote the measurement set at time step $k$ as $Z_{k}=\left\{ z_{k}^{1},...,z_{k}^{m_{k}}\right\} $.
We refer to measurement $z_{k}^{j}$ using the pair $\left(k,j\right)$
and the set of all such measurement pairs up to and including time
step $k$ is denoted by $\mathcal{M}_{k}$. Then, a local hypothesis
$a^{i}$ for $i=\left\{ 0,1,...,n_{k|k'}\right\} $ has an associated
set of measurement pairs denoted as $\mathcal{M}_{k}^{i,a^{i}}\subseteq\mathcal{M}_{k}$.
 The set $\mathcal{A}_{k|k'}$ of all global hypotheses meets
\begin{align*}
\mathcal{A}_{k|k'}= & \left\{ \left(a^{0},a^{1},...,a^{n_{k|k'}}\right):a^{i}\in\left\{ 1,...,h_{k|k'}^{i}\right\} \,\forall i,\right.\\
 & \left.\bigcup_{i=0}^{n_{k|k'}}\mathcal{M}_{k'}^{i,a^{i}}=\mathcal{M}_{k'},\mathcal{M}_{k'}^{i,a^{i}}\cap\mathcal{M}_{k'}^{j,a^{j}}=\emptyset,\,\forall i\neq j\right\} .
\end{align*}
That is, each measurement must be assigned to a local hypothesis,
and there cannot be more than one local hypothesis with the same measurement.
In this paper, we construct the (trees of) local hypotheses recursively,
and we allow for more than one measurement to be associated to the
same local hypothesis at the same time step, i.e., $\mathcal{M}_{k}^{i,a^{i}}$
may contain zero, one or more measurements. At time step zero, the
filter is initiated with $n_{0|0}=0$, $w_{0|0}^{0,1}=1$, $h_{0|0}^{0}=1$,
and $\mathcal{M}_{0}^{0,1}=\emptyset$. 

\section{General PMBM filter\label{sec:PMBM-filter-update}}

This section presents the PMBM filter for general target-generated
measurements and arbitrary clutter, with models described in Section
\ref{subsec:Models}. We consider the standard dynamic model so the
prediction step is the standard PMBM prediction step \cite{Williams15b,Angel18_b}.
The update is presented in Section \ref{subsec:General-PMBM-update}.
We provide a discussion and extension of the result to other filter
variants in Sections \ref{subsec:Discussion} and \ref{subsec:Extension-to-MBM,}.
The PMBM for point targets and arbitrary clutter is explained in Section
\ref{subsec:PMBM-point_targets-arbitrary_clutter}. Finally, an analysis
of the number of global hypotheses for different PMBM filters is provided
in Section \ref{subsec:Number-of-global-hypotheses}.

\subsection{General PMBM update\label{subsec:General-PMBM-update}}

Given two real-valued functions $a\left(\cdot\right)$ and $b\left(\cdot\right)$
on the target space, we denote their inner product as
\begin{align}
\left\langle a,b\right\rangle  & =\int a\left(x\right)b\left(x\right)dx.
\end{align}
Then, the update of the predicted PMBM $f_{k|k-1}\left(\cdot\right)$
with $Z_{k}$ is given in the following theorem.
\begin{thm}
\label{thm:PMBM_update_arbitrary_clutter}Assume the predicted density
$f_{k|k-1}\left(\cdot\right)$ is a PMBM of the form (\ref{eq:PMBM}).
Then, the updated density $f_{k|k}\left(\cdot\right)$ with set $Z_{k}=\left\{ z_{k}^{1},...,z_{k}^{m_{k}}\right\} $
is a PMBM with the following parameters. The number of Bernoulli components
is $n_{k|k}=n_{k|k-1}+2^{m_{k}}-1$. The intensity of the PPP is
\begin{align}
\lambda_{k|k}\left(x\right) & =f\left(\emptyset|x\right)\lambda_{k|k-1}\left(x\right).\label{eq:updated_PPP}
\end{align}

Let $Z_{k}^{1},...,Z_{k}^{2^{m_{k}}-1}$ be the non-empty subsets
of $Z_{k}$. The updated number of local clutter hypotheses is $h_{k|k}^{0}=2^{m_{k}}h_{k|k-1}^{0}$
such that a new local clutter hypothesis is included for each previous
local clutter hypothesis and either a misdetection or an update with
a non-empty subset of $Z_{k}$. The updated local clutter hypotheses
with no clutter at time step $k$, $a^{0}\in\left\{ 1,...,h_{k|k-1}^{0}\right\} $,
have parameters $\mathcal{M}_{k}^{0,a^{0}}=\mathcal{M}_{k-1}^{0,a^{0}}$,
\begin{align}
w_{k|k}^{0,a^{0}} & =w_{k|k-1}^{0,a^{0}}c\left(\emptyset\right).\label{eq:clutter_no_measurement_update}
\end{align}
For a previous local clutter hypothesis $\widetilde{a}^{0}\in\left\{ 1,...,h_{k|k-1}^{0}\right\} $
in the predicted density, the new local clutter hypothesis generated
by a set $Z_{k}^{j}$ has $a^{0}=\widetilde{a}^{0}+h_{k|k-1}^{0}j$,
\begin{align}
\mathcal{M}_{k}^{0,a^{0}} & =\mathcal{M}_{k-1}^{0,\widetilde{a}^{0}}\cup\left\{ \left(k,p\right):z_{k}^{p}\in Z_{k}^{j}\right\} ,\\
w_{k|k}^{0,a^{0}} & =w_{k|k-1}^{0,a^{0}}c\left(Z_{k}^{j}\right).\label{eq:clutter_measurement_update}
\end{align}

For Bernoulli components continuing from previous time steps $i\in\left\{ 1,...,n_{k|k-1}\right\} $,
a new local hypothesis is included for each previous local hypothesis
and either a misdetection or an update with a non-empty subset of
$Z_{k}$. The updated number of local hypotheses is $h_{k|k}^{i}=2^{m_{k}}h_{k|k-1}^{i}$.
For missed detection hypotheses, $i\in\left\{ 1,...,n_{k|k-1}\right\} $,
$a^{i}\in\left\{ 1,...,h_{k|k-1}^{i}\right\} $:
\begin{align}
\mathcal{M}_{k}^{i,a^{i}} & =\mathcal{M}_{k-1}^{i,a^{i}},\label{eq:Miss_measurement}\\
l_{k|k}^{i,a^{i},\emptyset} & =\big\langle p_{k|k-1}^{i,a^{i}},f\left(\emptyset|\cdot\right)\big\rangle,\label{eq:Miss_likelihood}\\
w_{k|k}^{i,a^{i}} & =w_{k|k-1}^{i,a^{i}}\left[1-r_{k|k-1}^{i,a^{i}}+r_{k|k-1}^{i,a^{i}}l_{k|k}^{i,a^{i},\emptyset}\right],\label{eq:Miss_weight}\\
r_{k|k}^{i,a^{i}} & =\frac{r_{k|k-1}^{i,a^{i}}l_{k|k}^{i,a^{i},\emptyset}}{1-r_{k|k-1}^{i,a^{i}}+r_{k|k-1}^{i,a^{i}}l_{k|k}^{i,a^{i},\emptyset}},\label{eq:Miss_existence}\\
p_{k|k}^{i,a^{i}}(x) & =\frac{f\left(\emptyset|x\right)p_{k|k-1}^{i,a^{i}}(x)}{l_{k|k}^{i,a^{i},\emptyset}}.\label{eq:Miss_density}
\end{align}

For a Bernoulli component $i\in\left\{ 1,...,n_{k|k-1}\right\} $
with a local hypothesis $\widetilde{a}^{i}\in\left\{ 1,...,h_{k|k-1}^{i}\right\} $
in the predicted density, the new local hypothesis generated by a
set $Z_{k}^{j}$ has $a^{i}=\widetilde{a}^{i}+h_{k|k-1}^{i}j$, $r_{k|k}^{i,a^{i}}=1$,
and
\begin{align}
\mathcal{M}_{k}^{i,a^{i}} & =\mathcal{M}_{k-1}^{i,\widetilde{a}^{i}}\cup\left\{ \left(k,p\right):z_{k}^{p}\in Z_{k}^{j}\right\} ,\\
l_{k|k}^{i,a^{i},Z_{k}^{j}} & =\bigg\langle p_{k|k-1}^{i,\widetilde{a}^{i}},f\left(Z_{k}^{j}|\cdot\right)\bigg\rangle,\\
w_{k|k}^{i,a^{i}} & =w_{k|k-1}^{i,\widetilde{a}^{i}}r_{k|k-1}^{i,\widetilde{a}^{i}}l_{k|k}^{i,a^{i},Z_{k}^{j}},\label{eq:update_weight}\\
p_{k|k}^{i,a^{i}}(x) & =\frac{f\left(Z_{k}^{j}|x\right)p_{k|k-1}^{i,\widetilde{a}^{i}}(x)}{l_{k|k}^{i,a^{i},Z_{k}^{j}}}.\label{eq:update_density}
\end{align}
For the new Bernoulli component initiated by subset $Z_{k}^{j}$,
whose index is $i=n_{k|k-1}+j$, we have two local hypotheses ($h_{k|k}^{i}=2$),
one corresponding to a non-existent Bernoulli density
\begin{equation}
\mathcal{M}_{k}^{i,1}=\emptyset,\;w_{k|k}^{i,1}=1,\;r_{k|k}^{i,1}=0,
\end{equation}
and the other with $r_{k|k}^{i,2}=1$, $w_{k|k}^{i,2}=l_{k|k}^{Z_{k}^{j}}$
and
\begin{align}
\mathcal{M}_{k}^{i,2} & =\left\{ \left(k,p\right):z_{k}^{p}\in Z_{k}^{j}\right\} ,\\
l_{k|k}^{Z_{k}^{j}} & =\bigg\langle\lambda_{k|k-1},f\left(Z_{k}^{j}|\cdot\right)\bigg\rangle,\label{eq:new_Bernoulli_likelihood}\\
p_{k|k}^{i,2}(x) & =\frac{f\left(Z_{k}^{j}|x\right)\lambda_{k|k-1}(x)}{l_{k|k}^{Z_{k}^{j}}}.\quad\square\label{eq:new_Bernoulli_density}
\end{align}
\end{thm}
Theorem \ref{thm:PMBM_update_arbitrary_clutter} is proved in Appendix
\ref{sec:AppendixA}. 

We can see that the updated PPP intensity in (\ref{eq:updated_PPP})
is the predicted intensity multiplied by the probability of receiving
no measurements, as in the PMBM filters for PPP clutter \cite{Williams15b,Angel18_b,Granstrom20,Angel21}.
Contrary to PMBM filters with PPP clutter, the general PMBM filter
extends local and global hypotheses to explicitly consider clutter
data associations, whose weights depend on the clutter density via
(\ref{eq:clutter_measurement_update}). As data associations for clutter
are separated from the target hypotheses, the probability of existence
of new Bernoulli components $r_{k|k}^{i,2}$ is equal to one, as in
the PMBM filter for unknown clutter rate in \cite{Si19}. This is
different from the PMBM filters for PPP clutter, where, for each separate
measurement, the hypotheses for clutter and the first detection of
a new target are merged into one. After merging, the probability of
existence of a new Bernoulli component depends on the clutter intensity,
and the number of global hypotheses for PPP clutter is lower, see
Section \ref{subsec:Set-of-global-hypotheses}. 

We also observe that each non-empty subset of $Z_{k}$ generates a
new Bernoulli, which has two local hypotheses. It is also possible
to instead reformulate the update to only generate $m_{k}$ new Bernoulli
components by increasing the number of local hypotheses of the new
Bernoulli components \cite[Sec. IV]{Xia22}. We proceed to illustrate
the differences between the hypotheses generated by this PMBM update,
and the update with PPP clutter with an example.
\begin{example}
Let us consider that the predicted density is a PPP and the measurement
set at time step 1 is $Z_{1}=\left\{ z_{1}^{1},z_{1}^{2}\right\} $.
The non-empty subsets are $Z_{1}^{1}=\left\{ z_{1}^{1}\right\} $,
$Z_{1}^{2}=\left\{ z_{1}^{2}\right\} $ and $Z_{1}^{3}=\left\{ z_{1}^{1},z_{1}^{2}\right\} $.
For general target-generated measurements (including extended targets)
with PPP clutter, we create 3 Bernoulli components, each initiated
by $Z_{1}^{1}$, $Z_{1}^{2}$ and $Z_{1}^{3}$, with two local hypotheses
\cite{Granstrom20,Angel21}. Therefore we have $\mathcal{M}_{1}^{1,2}=\left\{ \left(1,1\right)\right\} $,
$\mathcal{M}_{1}^{2,2}=\left\{ \left(1,2\right)\right\} $, $\mathcal{M}_{1}^{3,2}=\left\{ \left(1,1\right),\left(1,2\right)\right\} $.
The number of global hypotheses is the number of partitions of $Z_{1}$,
which is 2.

For general target-generated measurements and arbitrary clutter, Theorem
\ref{thm:PMBM_update_arbitrary_clutter} also creates 3 Bernoulli
components, each initiated by $Z_{1}^{1}$, $Z_{1}^{2}$ and $Z_{1}^{3}$,
with two local hypotheses. The difference now is that $\emptyset$,
$Z_{1}^{1}$, $Z_{1}^{2}$ and $Z_{1}^{3}$ can also be associated
to clutter, and each of these subsets generates a local hypothesis
for clutter, which has then 4 local hypotheses. That is, $\mathcal{M}_{1}^{0,1}=\emptyset$,
$\mathcal{M}_{1}^{0,2}=\left\{ \left(1,1\right)\right\} $, $\mathcal{M}_{1}^{0,3}=\left\{ \left(1,2\right)\right\} $,
and $\mathcal{M}_{1}^{0,4}=\left\{ \left(1,1\right),\left(1,2\right)\right\} $.
The number of global hypotheses is 5. $\diamondsuit$
\end{example}

\subsection{Discussion\label{subsec:Discussion}}

In this subsection, we discuss the spooky action at a distance that
may appear when we deal with arbitrary clutter, and an alternative
choice of hypotheses when the clutter density has some additional
structure. 

\subsubsection{Spooky action at a distance\label{subsec:Spooky-action}}

If there is one measurement far from all previous  targets represented
in the PMBM, the PMBM filters with PPP clutter generate a Bernoulli
component that appears in all global hypotheses and whose probability
of existence does not depend on the other measurements that have been
received in far away areas. That is, the probability of existence
of a newly detected isolated target only depends on the local situation. 

For arbitrary clutter PMBM filters, the new Bernoulli component has
probability of existence one in some, but not all, global hypotheses.
The weights of these global hypotheses depend on the clutter density
evaluated for all measurements. Therefore, the probability of existence
of the new target (considering all global hypotheses) can  depend
on all measurements, even if they are far away, a phenomenon usually
referred to as spooky action at a distance \cite{Franken09}. This
effect is illustrated the following example. 
\begin{example}
Let us consider the update of a PPP, which results in a PMB density,
with two measurements $\{z_{1}^{k},z_{2}^{k}\}$ that are far away
from each other. Then, a single target cannot generate both measurements,
and the probability of this event is zero. If clutter is PPP, the
probability of existence of a Bernoulli component generated by $z_{1}^{k}$
is \cite{Angel21}
\begin{align}
r_{k|k} & =\frac{l_{k|k}^{\left\{ z_{1}^{k}\right\} }}{\lambda^{C}\left(z_{1}^{k}\right)+l_{k|k}^{\left\{ z_{1}^{k}\right\} }}\label{eq:existence_PPP}
\end{align}
where $l_{k|k}^{\left\{ z_{1}^{k}\right\} }$ is given by (\ref{eq:new_Bernoulli_likelihood})
and $\lambda^{C}\left(\cdot\right)$ is the clutter intensity. The
probability of existence $r_{k|k}$ is independent of $z_{2}^{k}$. 

If clutter is arbitrary, there are four global hypotheses with non-negligible
weights, representing the events that $z_{1}^{k}$ and $z_{2}^{k}$
correspond to either a newly detected target or clutter. The probability
of existence, considering all global hypotheses, of the target initiated
by measurement $z_{1}^{k}$ is
\begin{align}
r_{k|k} & =\frac{l_{k|k}^{\left\{ z_{1}^{k}\right\} }\left(c\left(\left\{ z_{2}^{k}\right\} \right)+l_{k|k}^{\left\{ z_{2}^{k}\right\} }c\left(\emptyset\right)\right)}{w},\label{eq:existence_non_PPP}\\
w & =l_{k|k}^{\left\{ z_{1}^{k}\right\} }\left(c\left(\left\{ z_{2}^{k}\right\} \right)+l_{k|k}^{\left\{ z_{2}^{k}\right\} }c\left(\emptyset\right)\right)\nonumber \\
 & \quad+c\left(\left\{ z_{1}^{k}\right\} \right)l_{k|k}^{\left\{ z_{2}^{k}\right\} }+c\left(\left\{ z_{1}^{k},z_{2}^{k}\right\} \right).
\end{align}
If $c\left(\cdot\right)$ is a PPP, then, (\ref{eq:existence_non_PPP})
simplifies to (\ref{eq:existence_PPP}), and the probability of existence
is independent of $z_{2}^{k}$. However, in general, $r_{k|k}$ depends
on $z_{2}^{k}$, which implies that a far-away measurement influences
the probability of existence of a newly detected target in another
area, resulting in a spooky action at a distance phenomenon. $\diamondsuit$
\end{example}
It should also be noted that it is possible to design $c\left(\cdot\right)$
to avoid spooky action by setting it as the union of different, independent
sources of clutter in non-overlapping areas. 

\subsubsection{Alternative choice of hypotheses}

Theorem \ref{thm:PMBM_update_arbitrary_clutter} proves that the update
of a PMBM density with general target-generated measurements and arbitrary
clutter is a PMBM and provides an expression of the update. Nevertheless,
if the clutter has additional structure, for example, it is a PMB,
i.e., clutter being the union of a conventional PPP clutter plus $n_{c}$
Bernoulli sources of clutter, it is possible to design an alternative
hypothesis structure with $n_{c}$ sources of clutter and integrate
the PPP clutter into the probability of existence of new Bernoulli
components. Therefore, the clutter structure can be exploited to design
PMBM filters tailored to specific clutter models. An example of the
PMBM update with clutter that is the union of PPP clutter and a finite
number of independent clutter sources is provided in Appendix \ref{sec:AppendixC}.

\subsection{Extension to MBM, MBM$_{01}$, PMB, and labelled filters\label{subsec:Extension-to-MBM,}}

Theorem \ref{thm:PMBM_update_arbitrary_clutter} also holds for a
predicted density of the form MBM or MBM$_{01}$ by setting $\lambda_{k|k-1}\left(\cdot\right)=0$.
Therefore, we obtain the MBM filter by applying the standard MBM prediction
step \cite[Sec. III.E]{Angel18_b}, which includes multi-Bernoulli
birth model, followed by Theorem \ref{thm:PMBM_update_arbitrary_clutter}
update. The MBM$_{01}$ filter is the same as the MBM filter with
the difference that the MBM$_{01}$ filter prediction step expands
each Bernoulli density into two hypotheses with deterministic target
existence \cite[Eq. (36)]{Angel18_b}. This results in an unnecessary
exponential increase in the number of global hypotheses in the prediction
step, so it is not recommended. 

It is also relevant to notice that the general PMBM update in Theorem
\ref{thm:PMBM_update_arbitrary_clutter} also includes Bernoulli densities
with deterministic existence (when updating an existing Bernoulli
density with a non-empty measurement set or initiating a new Bernoulli
component). Nevertheless, these Bernoulli densities with deterministic
existence are required to deal with arbitrary clutter. In contrast,
the MBM$_{01}$ filter creates the deterministic Bernoulli densities
in the prediction, which is not necessary.

Both MBM and MBM$_{01}$ can be directly extended to include deterministic,
fixed labels assigned to each Bernoulli density of the birth model
without changing the filtering recursion \cite[Sec. IV.V]{Angel18_b}.
This procedure yields the labelled MBM and labelled MBM$_{01}$ filters.
The $\delta$-GLMB filtering recursion is analogous to the labelled
MBM$_{01}$ filtering recursion, so this procedure provides the $\delta$-GLMB
filter for general target-generated measurements and arbitrary clutter. 

PMB (and multi-Bernoulli) filters for arbitrary clutter can be obtained
by projecting the updated PMBM into a PMB using KLD minimisation \cite{Williams15b,Angel20_e,Williams15,Xia22}. 

\subsection{PMBM filter for point targets and arbitrary clutter\label{subsec:PMBM-point_targets-arbitrary_clutter}}

The equations in Theorem \ref{thm:PMBM_update_arbitrary_clutter}
are also valid for point targets with arbitrary clutter. In this setting,
the target-generated measurement density is given by (\ref{eq:standard_point_measurement}).
With the definitions of local and global hypotheses in Section \ref{subsec:Set-of-global-hypotheses},
many hypotheses associate more than one measurement to the same Bernoulli
component at the same time. However, according to (\ref{eq:standard_point_measurement}),
the weights of these hypotheses will be zero. Therefore, we can directly
exclude these hypotheses from the set $\mathcal{A}_{k|k}$ of global
hypotheses, by restricting the cardinality of $\mathcal{M}_{k}^{i,a^{i}}$
to zero or one for $i>0$. This approach, combined with the corresponding
prediction step, provides the PMBM, MBM, MBM$_{01}$ ($\delta$-GLMB)
filtering recursions for point targets and arbitrary clutter.

\subsection{Number of global hypotheses\label{subsec:Number-of-global-hypotheses}}

In this section, we calculate the number of global hypotheses of PMBM
filters to compare: 1) PPP clutter versus arbitrary clutter, and 2)
point targets versus the general target-generated measurement model.
This analysis helps us understand the differences in the hypothesis
structures of the PMBM filters and provides a measure of the complexity
of the corresponding data association problems.  To simplify the
analysis, we first consider the update of a PPP prior in Section \ref{subsec:Number_PPP_prior},
and then the update of a PMB prior in Section \ref{subsec:Number_PMB_prior}. 

\subsubsection{Update of a PPP prior\label{subsec:Number_PPP_prior}}

We calculate the number of global hypotheses when we update a PPP
prior. In the PMBM filter for point targets and PPP clutter, the number
of global hypotheses after updating a PPP prior is 1.  With a general
target-generated measurement model and PPP clutter, the number of
global hypotheses is the number of partitions of the measurement set,
which is the Bell number $B_{m_{k}}$ \cite[App. D]{Mahler_book14}. 

In the PMBM filter with point-target measurement model and arbitrary
clutter, the number of global hypotheses is $2^{m_{k}}$, representing
the events that each measurement is either a newly detected target
or clutter. With general target-generated measurements and arbitrary
clutter, the number of global hypotheses is
\begin{align}
 & \sum_{c=0}^{m_{k}}\left(\begin{array}{c}
m_{k}\\
c
\end{array}\right)B_{m_{k}-c}\label{eq:Number_global_hyp_gen_arb_PPP}
\end{align}
which is equal to $B_{m_{k}+1}$ applying \cite[Eq. (D.4)]{Mahler_book14}.
That is, in (\ref{eq:Number_global_hyp_gen_arb_PPP}), we go through
all possible numbers $c$ of clutter measurements. Then, the binomial
coefficient indicates the number of subsets with $c$ clutter measurements
and $B_{m_{k}-c}$ indicates the number of partitions of the rest
of the measurements, each generating a new global hypothesis. 

\subsubsection{Update of a Poisson multi-Bernoulli\label{subsec:Number_PMB_prior}}

We calculate the number of global hypotheses generated in the update
of the $j$-th global hypothesis of a prior PMBM, which represents
a PMB. To calculate the total number of global hypotheses, we sum
these results for all global hypotheses. The number of Bernoulli components
in the PMB with existence probability higher than zero is denoted
by $n^{j}$. 

For point targets and PPP clutter, the number of global hypotheses
is
\begin{align}
N_{p,p}\left(n^{j},m_{k}\right) & =\sum_{p=0}^{\min\left(n^{j},m_{k}\right)}p!\left(\begin{array}{c}
m_{k}\\
p
\end{array}\right)\left(\begin{array}{c}
n^{j}\\
p
\end{array}\right).\label{eq:N_pp}
\end{align}
That is, we go through all possible numbers $p$ of detected targets
among the ones that are represented by the Bernoulli components, the
number of ways of selecting $p$ measurements and $p$ targets, and
the number of permutations of $p$ elements (ways to associate $p$
measurements to $p$ targets). It can also be noticed that $m_{k}-p$
is the number of detections that are associated to either clutter
or newly detected targets.

For general target-generated measurements and PPP clutter, the number
of global hypotheses is \cite[Sec. V]{Granstrom18b}
\begin{align}
N_{g,p}\left(n^{j},m_{k}\right) & =\sum_{l=0}^{m_{k}}\left\{ \begin{array}{c}
m_{k}\\
l
\end{array}\right\} N_{p,p}\left(n^{j},l\right)\label{eq:N_gp}
\end{align}
where the factor in the brackets is the Stirling number of the second
kind, which indicates the number of ways of partitioning a set with
$m_{k}$ elements into $l$ non-empty cells. Starting the sum in (\ref{eq:N_gp})
with $l=0$ enables us to count one partition when $m_{k}=0$. 

For point targets and arbitrary clutter, the number of global hypotheses
is
\begin{align}
N_{p,a}\left(n^{j},m_{k}\right) & =\sum_{c=0}^{m_{k}}\left(\begin{array}{c}
m_{k}\\
c
\end{array}\right)N_{p,p}\left(n^{j},m_{k}-c\right).
\end{align}
That is, we first go through all possible clutter hypotheses with
$c$ clutter measurements. The remaining $m_{k}-c$ measurements can
be assigned either to previously detected targets or to new targets,
with at most one measurement per target. 

For general target-generated measurements and arbitrary clutter, the
number of global hypotheses is
\begin{align}
N_{g,a}\left(n^{j},m_{k}\right) & =\sum_{c=0}^{m_{k}}\left(\begin{array}{c}
m_{k}\\
c
\end{array}\right)N_{g,p}\left(n^{j},m_{k}-c\right).
\end{align}
That is, we first go through all possible clutter hypotheses with
$c$ clutter measurements. The remaining $m_{k}-c$ measurements are
assigned either to clutter or to new targets, following (\ref{eq:N_gp}). 

We show the number of global hypotheses of the filters after updating
a PMB prior for different number of measurements in Table \ref{tab:Number-global-hypotheses-PMB_prior},
where the case $n^{j}=0$ corresponds to a PPP prior. It can be observed
that considering general target-generated and arbitrary clutter significantly
increases the number of global hypotheses compared to point targets
and PPP clutter. 

\begin{table*}
\caption{\label{tab:Number-global-hypotheses-PMB_prior}Number of PMBM global
hypotheses after updating a PMB prior with $n^{j}$ targets and $m_{k}$
measurements}

\begin{centering}
\par\end{centering}
\centering{}%
\begin{tabular}{c|c|ccc|ccc|ccc|ccc}
\hline 
\multicolumn{2}{c|}{Target} &
\multicolumn{6}{c|}{Point} &
\multicolumn{6}{c}{General}\tabularnewline
\hline 
\multicolumn{2}{c|}{Clutter} &
\multicolumn{3}{c|}{PPP} &
\multicolumn{3}{c|}{Arbitrary} &
\multicolumn{3}{c|}{PPP} &
\multicolumn{3}{c}{Arbitrary}\tabularnewline
\hline 
\multicolumn{2}{c|}{$n^{j}$} &
0 &
1 &
4 &
0 &
1 &
4 &
0 &
1 &
4 &
0 &
1 &
4\tabularnewline
\hline 
\multirow{6}{*}{$m_{k}$} &
1 &
1 &
2 &
5 &
2 &
3 &
6 &
1 &
2 &
5 &
2 &
3 &
6\tabularnewline
 & 2 &
1 &
3 &
21 &
4 &
8 &
32 &
2 &
5 &
26 &
5 &
10 &
37\tabularnewline
 & 3 &
1 &
4 &
73 &
8 &
20 &
152 &
5 &
15 &
141 &
15 &
37 &
235\tabularnewline
 & 4 &
1 &
5 &
209 &
16 &
48 &
648 &
15 &
52 &
799 &
52 &
151 &
1540\tabularnewline
 & 5 &
1 &
6 &
501 &
32 &
112 &
2,512 &
52 &
203 &
4,736 &
203 &
674 &
10,427\tabularnewline
 & 10 &
1 &
11 &
8,501 &
1,024 &
6,144 &
850,944 &
115,975 &
678,570 &
67,310,847 &
678,570 &
3,535,027 &
247,126,450\tabularnewline
\hline 
\end{tabular}
\end{table*}

\section{Data associations in point-target PMBM filter with arbitrary clutter
via Gibbs sampling\label{sec:Data-associations-Gibbs}}

In this section, we develop a tractable PMBM filter for point targets
and arbitrary clutter. A key challenge in this context is to handle
the intractable growth in the number of data association hypotheses.
For PPP clutter, we can use Murty's algorithm to select the $K$-best
updated global hypotheses generated from a single predicted global
hypothesis, representing a PMB \cite{Murty68,Angel18_b}, and then
prune the other hypotheses. However, for arbitrary clutter, the data
association problem is no longer an assignment problem \cite{Crouse16}
and thus we cannot use Murty's algorithm. In this section, we explain
how to use Gibbs sampling to select $K$-good updated global hypotheses
from a predicted global hypothesis \cite{Hue02,Hue02b,Vo17,Granstrom18b}.
That is, Gibbs sampling does not necessarily find the $K$-best global
hypotheses, but it samples global hypotheses according to their weights,
which implies that global hypotheses with higher weights are more
likely to be selected. 

\subsection{Notation}

To simplify notation in this section, we drop time indices in the
number of measurements $m$ and the measurement set $\left\{ z^{1},...,z^{m}\right\} $.
In addition, the Bernoulli densities for the predicted global hypothesis,
representing a PMB, have parameters $\left(r^{i},p^{i}(\cdot)\right)$
for $i\in\left\{ 1,...,n_{k|k-1}\right\} $. 

Then, we recall that for the point target model, after the update,
we have $n_{k|k}=n_{k|k-1}+m$ Bernoulli components, which includes
$n_{k|k-1}$ predicted Bernoulli components and the new $m$ Bernoulli
components, one generated by each measurement. The data associations
can be represented as a vector $\gamma_{1:m}=\left(\gamma_{1},...,\gamma_{m}\right)$
where $\gamma_{j}\in\{0,1,...,n_{k|k}\}$ is the Bernoulli component
index associated to the $j$-th measurement and $\gamma_{i}\neq\gamma_{j}$
for all $i\neq j$ such that $\gamma_{i}>0$ and $\gamma_{j}>0$.
If $\gamma_{j}=0$, it means that the $j$-th measurement is clutter.
The set of vectors $\gamma_{1:m}$ that meet this property is denoted
by $\Gamma$.

We also use $1_{\Gamma}\left(\cdot\right)$ to denote the indicator
function of set $\Gamma$: $1_{\Gamma}\left(\gamma_{1:m}\right)=1$
if $\gamma_{1:m}\in\Gamma$, and $1_{\Gamma}\left(\gamma_{1:m}\right)=0$
otherwise. 

\subsection{Arbitrary clutter density}

The probability of data association $\gamma_{1:m}$, without accounting
for the weight of the predicted global hypothesis, is
\begin{align}
p\left(\gamma_{1:m}\right) & \propto1_{\Gamma}\left(\gamma_{1:m}\right)c\left(Z_{c}\left(\gamma_{1:m}\right)\right)\prod_{j=1:\gamma_{j}>0}^{m}\eta^{j}\left(\gamma_{j}\right)\label{eq:density_gamma_1_m}
\end{align}
where the measurement set corresponding to clutter is
\begin{align}
Z_{c}\left(\gamma_{1:m}\right) & =\left\{ z^{j}:\gamma_{j}=0\right\} 
\end{align}
and
\begin{align}
 & \eta^{j}\left(\gamma_{j}\right)\nonumber \\
 & \;=\begin{cases}
\frac{r^{\gamma_{j}}\int p^{D}\left(x\right)l\left(z^{j}|x\right)p^{\gamma_{j}}(x)dx}{1-r^{\gamma_{j}}\left\langle p^{\gamma_{j}},p^{D}\right\rangle } & \gamma_{j}\leq n_{k|k-1}\\
\int p^{D}\left(x\right)l\left(z^{j}|x\right)\lambda_{k|k-1}\left(x\right)dx & n_{k|k-1}<\gamma_{j}\leq n_{k|k}.
\end{cases}\label{eq:eta_data_associations}
\end{align}
In (\ref{eq:eta_data_associations}), the second line corresponds
to the weight of new Bernoulli components, see (\ref{eq:new_Bernoulli_likelihood}).
The first line corresponds to the weight of the predicted Bernoulli
components with a detection $z^{j}$, see (\ref{eq:update_weight}),
divided by the misdetection weight, see (\ref{eq:Miss_weight}), both
using the point target measurement model (\ref{eq:standard_point_measurement}).

To obtain samples from density (\ref{eq:density_gamma_1_m}) using
Gibbs sampling, we calculate the conditional density \cite{Liu_book01}
\begin{align}
p\left(\gamma_{q}|\gamma_{1:q-1},\gamma_{q+1:m}\right) & \propto p\left(\gamma_{1:m}\right)\\
 & \propto1_{\Gamma}\left(\gamma_{1:m}\right)c\left(Z_{c}\left(\gamma_{1:m}\right)\right)\nonumber \\
 & \quad\times\prod_{j=1:\gamma_{j}>0}^{m}\eta^{j}\left(\gamma_{j}\right)\label{eq:conditional_assignment}
\end{align}
where $q\in\{1,...,m\}$. As $\gamma_{1:q-1},\gamma_{q+1:m}$ are
constants in (\ref{eq:conditional_assignment}), $l^{j}\left(\gamma_{j}\right)$
for $j\neq q$ are constants too, and we can write
\begin{align}
 & p\left(\gamma_{q}|\gamma_{1:q-1},\gamma_{q+1:m}\right)\nonumber \\
 & \propto\begin{cases}
1_{\Gamma}\left(\gamma_{1:m}\right)c\left(Z_{c}\left(\gamma_{1:m}\right)\right)\eta^{q}\left(\gamma_{q}\right) & \gamma_{q}>0\\
c\left(Z_{c}\left(\gamma_{1:m}\right)\right) & \gamma_{q}=0
\end{cases}\label{eq:conditional_assignment-split}
\end{align}
where we have used that $\left(\gamma_{1:q-1},0,\gamma_{q+1:m}\right)\in\Gamma$.
Equation (\ref{eq:conditional_assignment-split}) can be evaluated
for all possible values $\gamma_{q}\in\{0,1,...,m\}$, and therefore,
we can sample from this categorical distribution. To perform Gibbs
sampling, we can start with $\gamma_{1:m}$ having all entries equal
to zero, and sample $\gamma_{q}$ from $q=1$ to $m$ for a total
number of $K$ times. Then, as in \cite{Vo17}, we remove the repeated
samples of data association hypotheses to yield the newly generated
data association hypotheses. In the implementations, we set $K=\left\lceil N_{h}\cdot w_{j}\right\rceil $,
where $N_{h}$ is the maximum number of global hypotheses. 

\subsection{Uniform IID cluster clutter}

In this subsection, we specify (\ref{eq:conditional_assignment-split})
for the case where the clutter $c\left(\cdot\right)$ follows an IID
cluster process \cite{Mahler_book14} whose the single-measurement
density $\breve{c}\left(\cdot\right)$ is uniform in the field-of
view $A$. That is,
\begin{align}
c\left(Z\right) & =|Z|!\rho_{c}\left(|Z|\right)\prod_{z\in Z}\breve{c}\left(z\right)\label{eq:IID_cluster_clutter}\\
\breve{c}\left(z\right) & =u_{A}\left(z\right)\label{eq:uniform_clutter_density}
\end{align}
where $\rho_{c}\left(|Z|\right)$ is the cardinality distribution
and $u_{A}\left(\cdot\right)$ is a uniform density in $A$. 

Let $m_{c}$ denote the number of clutter measurements in $\gamma_{1:q-1},\gamma_{q+1:m}$,
i.e.,
\begin{align}
m_{c} & =\sum_{j=1:j\neq q}^{m}\delta_{0}\left[\gamma_{j}\right]
\end{align}
where $\delta_{j}[\cdot]$ is the Kronecker delta  at $j$. Then,
considering that all measurements are in the field-of-view, (\ref{eq:conditional_assignment-split})
becomes
\begin{align}
 & p\left(\gamma_{q}|\gamma_{1:q},\gamma_{q+1:m}\right)\nonumber \\
 & \propto\begin{cases}
1_{\Gamma}\left(\gamma_{1:m}\right)\eta^{q}\left(\gamma_{q}\right)\rho_{c}\left(m_{c}\right)m_{c}!\frac{1}{|A|^{m_{c}}} & \gamma_{q}>0\\
\rho_{c}\left(m_{c}+1\right)\left(m_{c}+1\right)!\frac{1}{|A|^{m_{c}+1}} & \gamma_{q}=0
\end{cases}\label{eq:conditional_assignment-uniform_clutter}
\end{align}
where $|A|$ is the area (or the hypervolume) of $A$.

\section{Simulation experiments\label{sec:Simulation-experiments}}

In this section, we evaluate the developed PMBM and PMB filters in
two scenarios\footnote{Matlab code is available at https://github.com/Agarciafernandez and
https://github.com/yuhsuansia.}. The main objective of these simulations is to study how filters
deal with non-standard sources of clutter. Section \ref{subsec:Point-targets-NB-clutter}
presents a scenario with point targets and IID cluster clutter with
a negative binomial cardinality distribution. Section \ref{subsec:Extended-targets-independent_clutter}
presents a scenario with extended targets and a number of independent
clutter sources of clutter.

\subsection{Point targets and negative binomial IID cluster clutter\label{subsec:Point-targets-NB-clutter}}

In this section, we consider point targets with clutter being an IID
cluster process with negative binomial (NB) cardinality distribution
\cite{Zhou15b}. We compare the following filters: the standard PMBM
and PMB filter (with PPP clutter assumption), and the arbitrary clutter
PMBM and PMB filters, referred to as A-PMBM and A-PMB. The rest of
the parameters are \cite{Angel18_b}: maximum number of global hypotheses
$N_{h}=5,000$, threshold for MBM pruning $10^{-4}$, threshold for
pruning the PPP weights $\Gamma_{p}=10^{-5}$, threshold for pruning
Bernoulli densities $\Gamma_{b}=10^{-5}$, and ellipsoidal gating
with threshold 20. These filters use Estimator 3 as it provides the
best performance among the standard PMBM estimators \cite[Sec. VI]{Angel18_b}
in this scenario. Estimator 3 obtains the global hypothesis with a
deterministic cardinality with highest weight and then reports the
mean of the targets in this hypothesis. 

Additionally, we have implemented two filters with multi-Bernoulli
birth and PPP clutter: the MBM filter \cite{Angel18_b,Angel19_e},
and the $\delta$-GLMB filter with joint prediction and update \cite{Vo17}.
The $\delta$-GLMB filter has been implemented with 7,000 global hypotheses
and pruning threshold $10^{-15}$. All the previous filters have
been implemented using Gibbs sampling to select relevant global hypotheses
in the update of each predicted global hypothesis. We have also implemented
a Gaussian mixture PHD filter for PPP clutter \cite{Mahler_book14,Vo06}
and a Gaussian mixture PHD filter for NB IID clutter in \cite{Schlangen16},
which we refer to as the NB-PHD filter. The PHD filters have parameters:
pruning threshold $10^{-5}$, merging threshold 0.1, and maximum number
of Gaussian components 30.

\subsubsection{Clutter density}

We consider an IID cluster clutter density with uniform single-measurement
density in the field of view, as in (\ref{eq:IID_cluster_clutter})
and (\ref{eq:uniform_clutter_density}). Its cardinality distribution
is NB, $\rho_{c}\left(m\right)=\mathrm{NB}\left(m;r,p\right)$, with
$r>0$, $p\in\left[0,1\right]$, and \cite{Zhou15b}
\begin{align}
\mathrm{NB}\left(m;r,p\right) & =\frac{\Gamma\left(r+m\right)}{\Gamma\left(r\right)\Gamma\left(m+1\right)}p^{r}\left(1-p\right)^{m}\label{eq:NB_density}
\end{align}
where $\Gamma\left(\cdot\right)$ is the gamma function. Its mean
and variance are
\begin{equation}
\mathrm{E}\left[m\right]=\frac{\left(1-p\right)r}{p},\:\mathrm{V}\left[m\right]=\frac{\left(1-p\right)r}{p^{2}}.\label{eq:NB_mean_variance}
\end{equation}
Therefore, if we denote $\mathrm{E}\left[m\right]=\overline{\lambda}^{C}$
and $\mathrm{V}\left[m\right]=a^{C}\overline{\lambda}^{C}$ with $a^{C}>1$,
using (\ref{eq:NB_density}), we can obtain parameters $r$ and $p$
based on $\overline{\lambda}^{C}$ and $a^{C}$ 
\begin{equation}
r=\frac{\overline{\lambda}^{C}}{a^{C}-1},\;p=\frac{1}{a^{C}}.
\end{equation}
Parameter $a^{C}$ indicates the over-dispersion w.r.t. the Poisson
distribution. In fact, the NB distribution results from integrating
out the rate of a Poisson distribution, with the rate having a gamma
distribution \cite{Zhou15b}, so it is always over-dispersed. This
means that for NB distribution its variance is always larger than
its mean, making it suitable for scenarios with high clutter variance.
For $a^{C}\rightarrow1$, the NB distribution tends to a Poisson distribution
with parameter $\overline{\lambda}^{C}$. 

\begin{figure}
\begin{centering}
\includegraphics[scale=0.6]{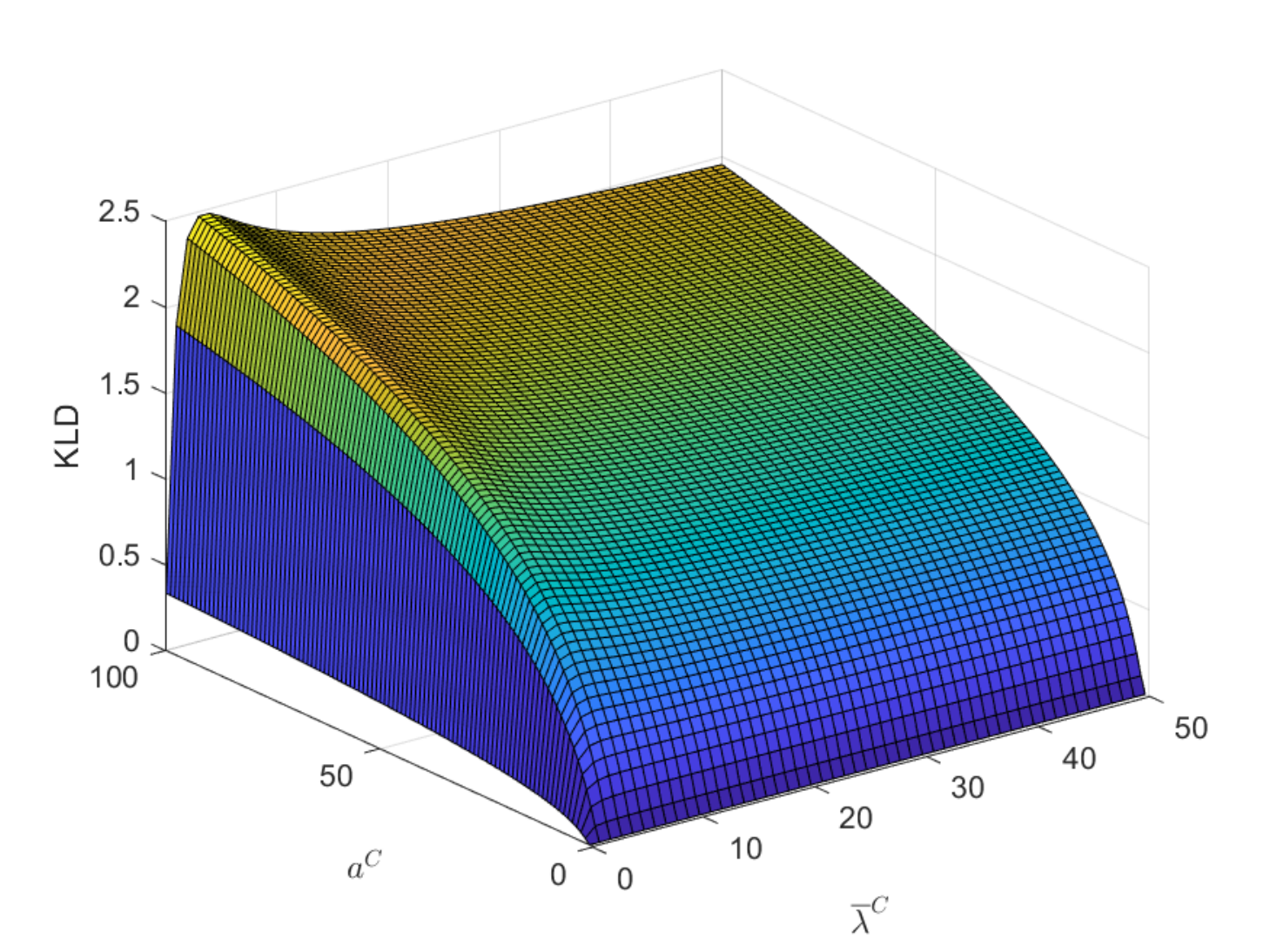}
\par\end{centering}
\caption{\label{fig:KLD-NB-Poisson}KLD $\mathrm{D}\left(\mathrm{P}||\mathrm{NB}\right)$
between Poisson and negative binomial distributions for different
values of $\overline{\lambda}^{C}$ and $a^{C}$. As $a^{C}$ increases,
the distributions become more different.}
\end{figure}

The KLD $\mathrm{D}\left(\mathrm{P}||\mathrm{NB}\right)$ between
the Poisson (P) and NB distributions with the same mean $\overline{\lambda}^{C}$
w.r.t. different values of $\overline{\lambda}^{C}$ and $a^{C}$
is shown in Figure \ref{fig:KLD-NB-Poisson}. We can see that, for
$a^{C}\rightarrow1$ and $\overline{\lambda}^{C}\rightarrow0$, the
distributions tend to be more similar, and they differ as these parameters
increase, mainly due to increments in $a^{C}$. This is important
as it implies that if the $\mathrm{NB}$ distribution is quite similar
to the Poisson, then, the PMBM filter with PPP clutter is expected
to outperform the PMBM filter with arbitrary clutter due to its improved
hypothesis structure, as pointed out in Section \ref{subsec:Number-of-global-hypotheses}.

\subsubsection{Simulation results}

The target state is $x=[p_{x},v_{x},p_{y},v_{y}]^{\top}$, which contains
its position and velocity. The dynamic model is a nearly-constant
velocity model \cite{Bar-Shalom_book01}
\begin{align*}
g\left(x_{k}|x_{k-1}\right) & =\mathcal{N}\left(x_{k};Fx_{k-1},Q\right)
\end{align*}
\[
F=I_{2}\otimes\begin{bmatrix}1 & T\\
0 & 1
\end{bmatrix},\,Q=qI_{2}\otimes\begin{bmatrix}T^{3}/3 & T^{2}/2\\
T^{2}/2 & T
\end{bmatrix},
\]
where $\otimes$ is the Kronecker product, $q=0.01\,\mathrm{m}^{2}/\mathrm{s}^{3}$,
the sampling time $T=1\,\mathrm{s}$, and $\mathcal{N}\left(x;\overline{x},P\right)$
represents a Gaussian density with mean $\overline{x}$ and covariance
matrix $P$ evaluated at $x$. The probability of survival is $p^{S}=0.99$.
The birth process is a PPP with a Gaussian intensity
\begin{align}
\lambda_{k}^{B}\left(x\right) & =w_{k}^{b}\mathcal{N}\left(x;\overline{x}_{k}^{b},P_{k}^{b}\right),\label{eq:PPP_Gaussian_birth}
\end{align}
where $\overline{x}_{k}^{b}=\left[150\,(\mathrm{m}),0\,(\mathrm{m/s}),150\,(\mathrm{m}),0\,(\mathrm{m/s})\right]^{\top}$,
$P_{k}^{b}=\mathrm{diag}\left(\left[50^{2}\,(\mathrm{m}^{2}),1\,(\mathrm{m}^{2}/\mathrm{s}^{2}),50^{2}\,(\mathrm{m}^{2}),1\,(\mathrm{m}^{2}/\mathrm{s}^{2})\right]\right)$
and the weight $w_{1}^{b}=5$ for $k=1$, and $w_{k}^{b}=0.1$ for
$k>1$. The weight $w_{k}^{b}$ represents the expected number of
new born targets at time step $k$ \cite{Mahler_book14}. 

\begin{figure}
\begin{centering}
\includegraphics[scale=0.6]{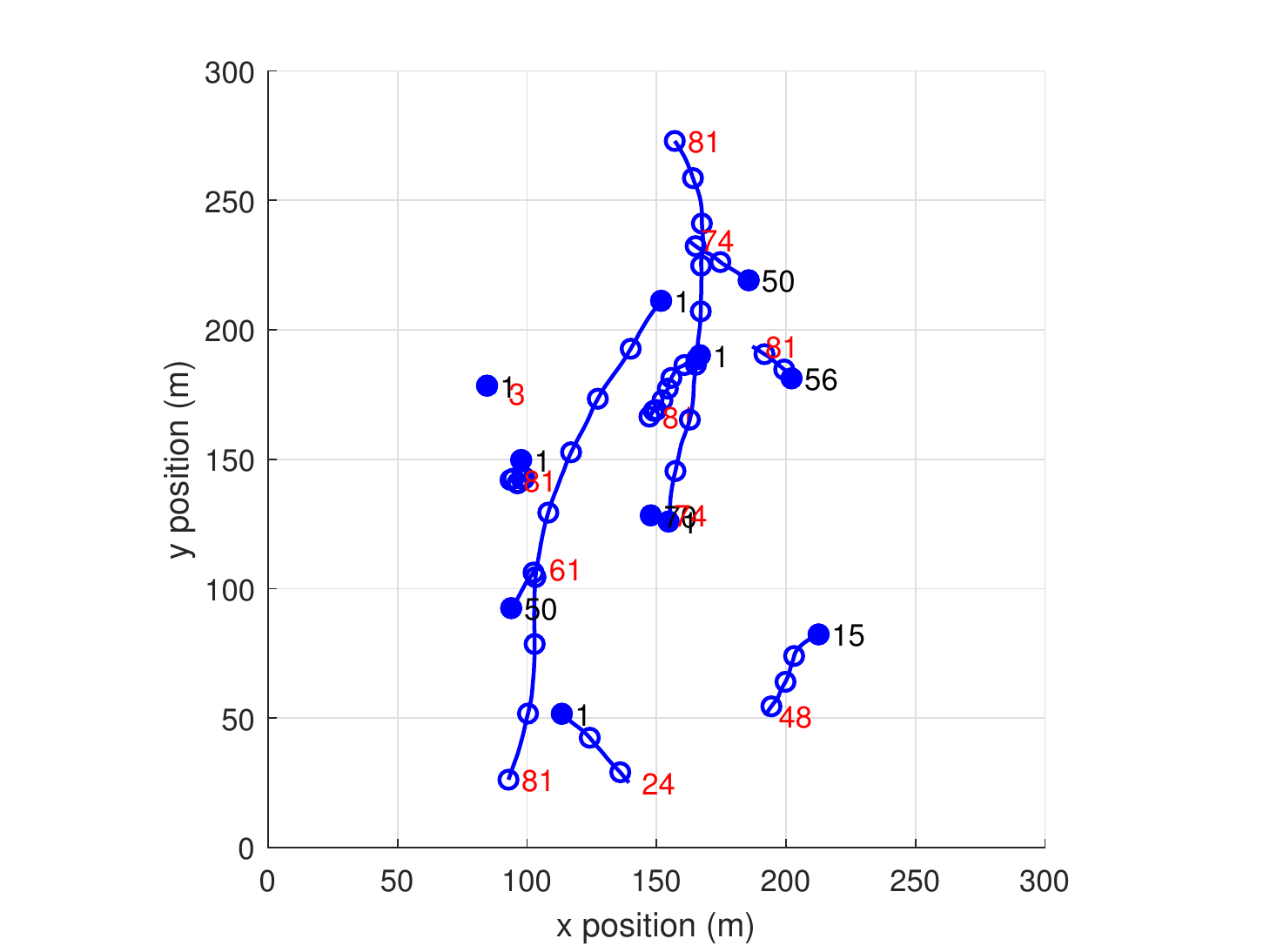}
\par\end{centering}
\caption{\label{fig:Ground-truth-point_targets}Ground truth set of trajectories
for the point target scenario. Initial trajectory positions are marked
with filled circles and their positions every ten time steps are marked
with a circle. The black numbers next to birth positions indicate
the time of birth and the red numbers the last time a target is alive.}
\end{figure}

The ground truth set of trajectories, shown in Figure \ref{fig:Ground-truth-point_targets},
is generated by sampling the above dynamic model for 81 time steps.
The target starting at time step 1 at around $[150,210]\,(\mathrm{m}),$
and the one starting at time step 50 at around $[95,95]\,(\mathrm{m})$
get in close proximity at around time step 52.

The filters with multi-Bernoulli birth (MBM and $\delta$-GLMB) use
9 Bernoulli birth components at time step 1 with probability of existence
5/9, mean $\overline{x}_{k}^{b}$ and covariance matrix $P_{k}^{b}$.
This choice approximately covers the support of the cardinality of
the PPP, and sets the same PHD for the multi-Bernoulli and PPP birth
models \cite{Mahler_book14}. For $k>1$, these filters use a Bernoulli
birth with probability of existence $w_{k}^{b}$, mean $\overline{x}_{k}^{b}$
and covariance matrix $P_{k}^{b}$. 

At each time step, the sensor measures positions with likelihood
\begin{align*}
l\left(z|x\right) & =\mathcal{N}\left(z;Hx,R\right)
\end{align*}
\[
H=I_{2}\otimes\begin{bmatrix}1 & 0\end{bmatrix},\,R=4I_{2}\,(\mathrm{m}^{2}),
\]
and a probability of detection $p^{D}=0.9$. Clutter is uniformly
distributed in the region of interest $A=[0,300]\times[0,300]\,(\mathrm{m}\times\mathrm{m})$
such that $\breve{c}\left(z\right)=u_{A}\left(z\right)$. The NB clutter
parameters are $a^{C}=20$ and $\overline{\lambda}^{C}=10$.

\begin{figure}
\begin{centering}
\includegraphics[scale=0.3]{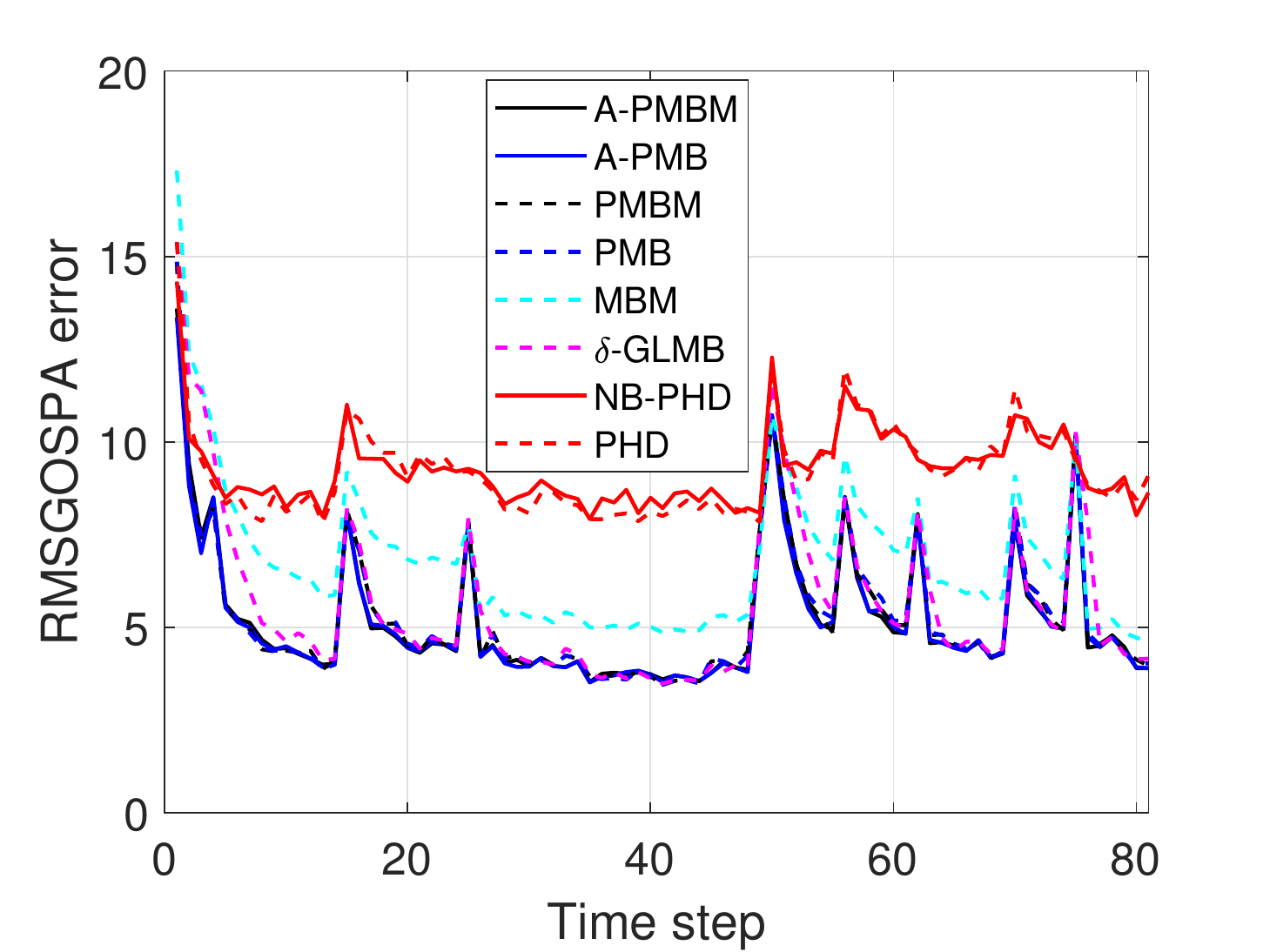}\includegraphics[scale=0.3]{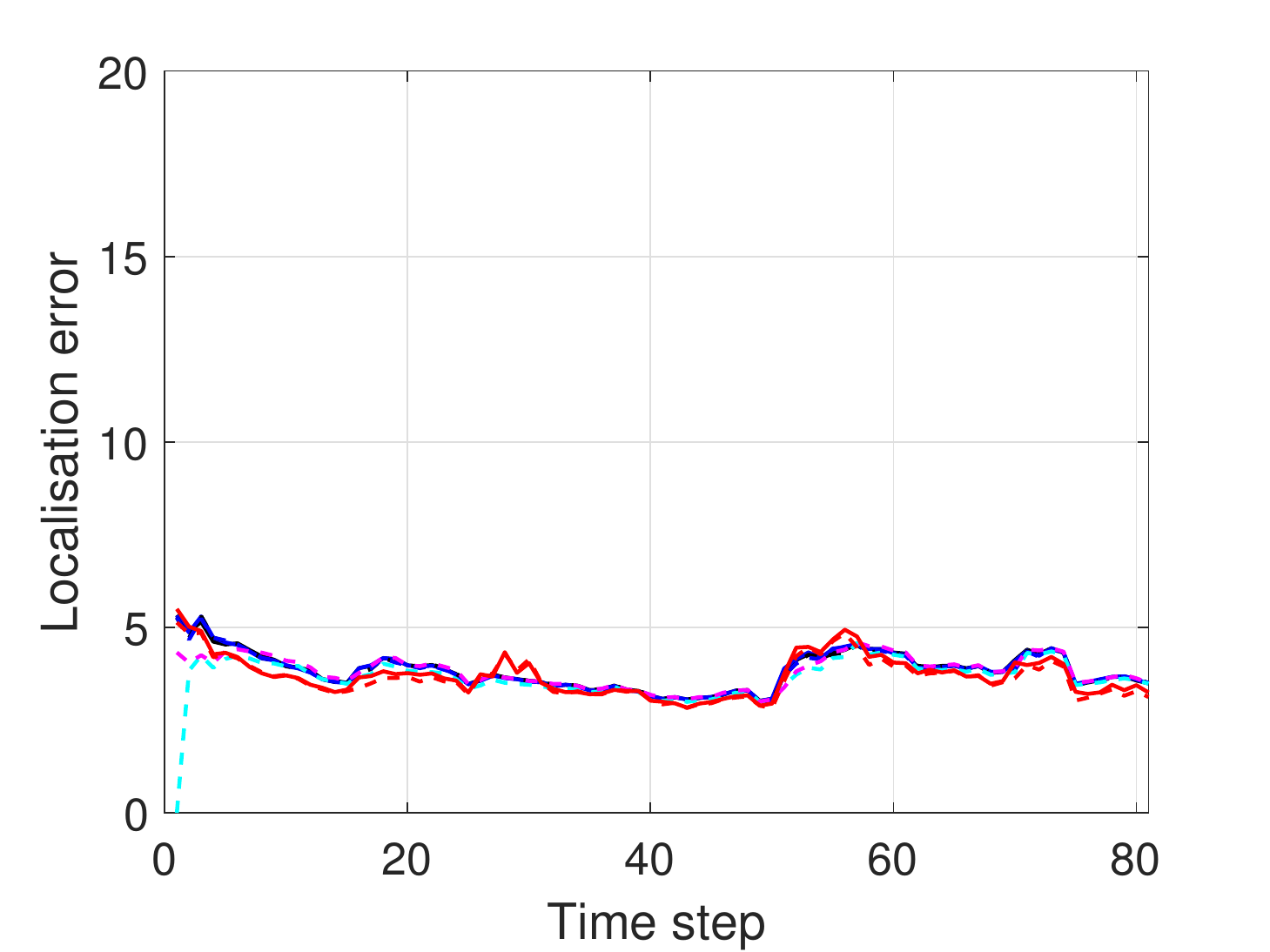}
\par\end{centering}
\begin{centering}
\includegraphics[scale=0.3]{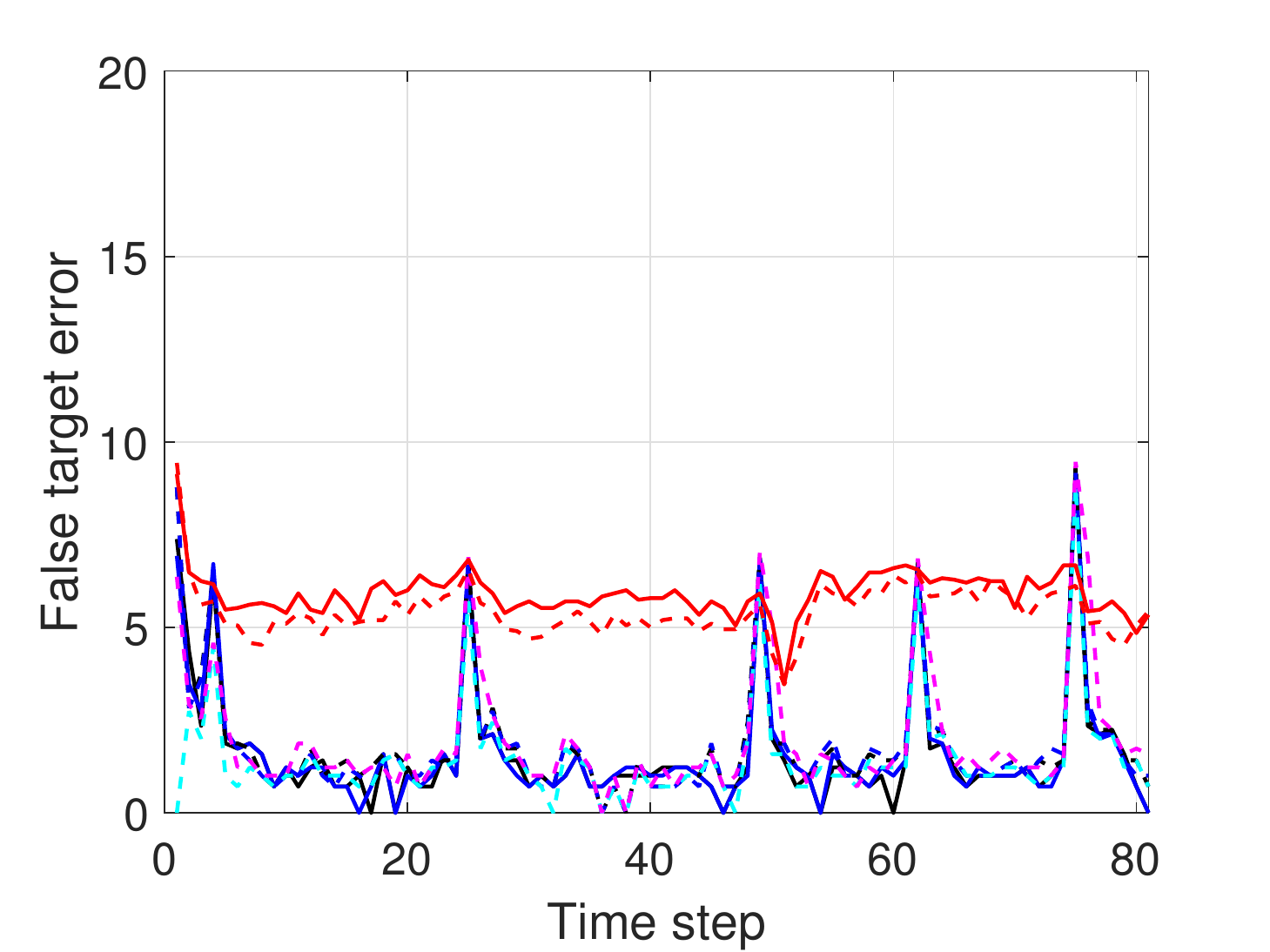}\includegraphics[scale=0.3]{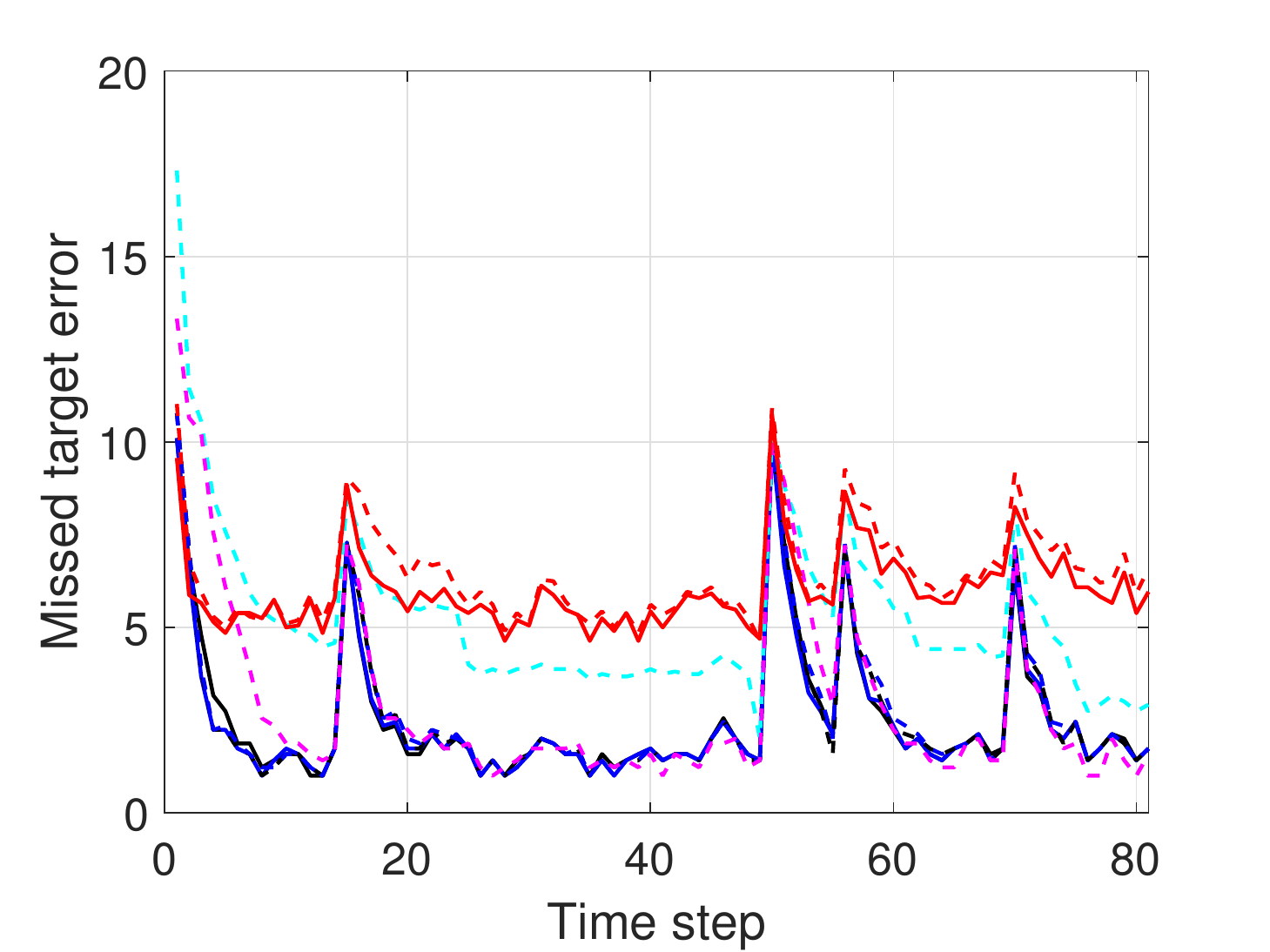}
\par\end{centering}
\caption{\label{fig:RMS-GOSPA-errors}RMS GOSPA errors ($\mathrm{m)}$ and
their decomposition against time for the point target scenario. }
\end{figure}

We evaluate the performances of the filters using Monte Carlo simulation
with 100 runs. We obtain the root mean square generalised optimal
subpattern assignment (RMS-GOSPA) metric error ($p=2$, $c=10\,\mathrm{m}$,
$\alpha=2$) for the position elements at each time step \cite{Rahmathullah17}.
The resulting GOSPA errors as well as their decomposition into localisation
error for properly detected targets, and costs for missed and false
targets are shown in Figure \ref{fig:RMS-GOSPA-errors}. The peaks
in Figure \ref{fig:RMS-GOSPA-errors} correspond to the time steps
in which targets are born or die, see Figure \ref{fig:Ground-truth-point_targets}.
These peaks arise as it usually takes at least 2 time steps to estimate
a new born target and also to stop reporting it when it dies. These
peaks happen for all filters, though they are more difficult to spot
on the PHD filters as they have a higher number of missed and false
targets throughout all time steps. A-PMBM and A-PMB filters are the
filters with best performance, closely followed by PMBM and PMB filters.
The $\delta$-GLMB filter provides higher errors, specially due to
missed targets. The MBM filter performs worse, and the PHD filters
have the worst performance, as they provide a less accurate approximation
to the posterior.

The computational times in seconds to run one Monte Carlo simulation
on an Intel core i5 laptop are: 43.4 (A-PMBM), 40.7 (A-PMB), 20.2
(PMBM), 14.7 (PMB), 20.8 (MBM), 45.6 ($\delta$-GLMB), 1.1 (NB-PHD)
and 1.1 (PHD). The PHD filters are the fastest algorithms, with the
lowest performance. This is to be expected as they propagate a PPP
through the filtering recursion so the structure of the posterior
and the data association problem is simplified at the cost of lower
performance. The A-PMBM and A-PMB filters have a higher computational
burden than PMBM and PMB filters due to the higher number of global
hypotheses and the extra calculations with negative binomial clutter.
The $\delta$-GLMB filter has a higher computational burden due to
its global hypothesis structure and implementation parameters \cite{Angel18_b}.

\begin{table}
\caption{\label{tab:RMS-GOSPA-errors_pd}RMS-GOSPA errors ($\mathrm{m)}$ across
time for different $p^{D}$}

\centering{}%
\begin{tabular}{c|cccc}
\hline 
$p^{D}$ &
0.95 &
0.9 &
0.8 &
0.7\tabularnewline
\hline 
A-PMBM &
5.18 &
5.53 &
6.17 &
6.85\tabularnewline
A-PMB &
\uline{5.12} &
\uline{5.50} &
\uline{6.16} &
\uline{6.82}\tabularnewline
PMBM &
5.25 &
5.65 &
6.34 &
7.06\tabularnewline
PMB &
5.26 &
5.67 &
6.38 &
7.10\tabularnewline
\hline 
MBM &
7.16 &
7.18 &
7.50 &
8.07\tabularnewline
$\delta$-GLMB &
5.74 &
6.10 &
6.74 &
7.60\tabularnewline
\hline 
NB-PHD &
6.86 &
9.31 &
11.56 &
13.28\tabularnewline
PHD &
7.12 &
9.28 &
12.04 &
13.72\tabularnewline
\hline 
\end{tabular}
\end{table}

The RMS-GOSPA errors considering all time steps in the simulation
for several values of $p^{D}$ are shown in Table \ref{tab:RMS-GOSPA-errors_pd}.
We can see that the filter with lowest error for all $p^{D}$ is the
A-PMB filter tightly followed by the A-PMBM filter. While A-PMBM is
the theoretical optimal filter, for a fixed number of global hypotheses
and a sub-optimal estimator, A-PMB can outperform A-PMBM, especially
in scenarios without challenging multi-target crossings. In particular,
at each A-PMBM update, new Bernoulli components have probability of
existence either 0 or 1, which requires a high number of global hypotheses
(see Section \ref{subsec:Number-of-global-hypotheses}) that are updated
separately at subsequent time steps. On the other hand, the A-PMB
filter projects all the 0-1 hypotheses from the same Bernoulli component
into a single Bernoulli density, providing a compact representation.
As expected, the higher the probability of detection, the lower the
error for all filters.

\begin{table*}
\caption{\label{tab:RMS-GOSPA-errors-across_clutter}RMS-GOSPA errors ($\mathrm{m)}$
across time for different NB clutter parameters}

\centering{}%
\begin{tabular}{ccccc|cccc|cccc|cccc}
\hline 
$a^{C}$ &
\multicolumn{4}{c|}{2} &
\multicolumn{4}{c|}{10} &
\multicolumn{4}{c|}{20} &
\multicolumn{4}{c}{40}\tabularnewline
\hline 
$\overline{\lambda}^{C}$ &
1 &
5 &
10 &
15 &
1 &
5 &
10 &
15 &
1 &
5 &
10 &
15 &
1 &
5 &
10 &
15\tabularnewline
\hline 
A-PMBM &
5.15 &
\uline{5.45} &
5.71 &
5.98 &
\uline{5.05} &
5.35 &
5.63 &
5.84 &
\uline{5.04} &
5.29 &
5.53 &
5.76 &
\uline{4.93} &
5.21 &
5.42 &
5.60\tabularnewline
A-PMB &
5.14 &
\uline{5.45} &
\uline{5.65} &
\uline{5.86} &
\uline{5.05} &
\uline{5.34} &
\uline{5.59} &
\uline{5.76} &
\uline{5.04} &
\uline{5.28} &
\uline{5.50} &
\uline{5.69} &
4.95 &
\uline{5.20} &
\uline{5.39} &
\uline{5.56}\tabularnewline
PMBM &
\uline{5.13} &
\uline{5.45} &
\uline{5.65} &
\uline{5.86} &
5.11 &
5.41 &
5.66 &
5.88 &
5.14 &
5.40 &
5.65 &
5.88 &
5.13 &
5.40 &
5.62 &
5.83\tabularnewline
PMB &
5.14 &
5.46 &
5.67 &
5.88 &
5.12 &
5.41 &
5.69 &
5.90 &
5.15 &
5.43 &
5.67 &
5.91 &
5.14 &
5.42 &
5.64 &
5.86\tabularnewline
\hline 
MBM &
5.27 &
6.23 &
6.32 &
6.40 &
5.22 &
6.61 &
6.82 &
6.85 &
5.27 &
6.91 &
7.12 &
7.18 &
5.26 &
7.09 &
7.38 &
7.61\tabularnewline
$\delta$-GLMB &
5.36 &
5.73 &
5.99 &
6.24 &
5.40 &
5.79 &
6.03 &
6.28 &
5.43 &
5.84 &
6.10 &
6.31 &
5.42 &
5.85 &
6.08 &
6.32\tabularnewline
\hline 
NB-PHD &
9.13 &
9.22 &
9.42 &
9.73 &
8.90 &
9.22 &
9.29 &
9.54 &
8.87 &
9,08 &
9.31 &
9.48 &
8.86 &
8.99 &
9.24 &
9.34\tabularnewline
PHD &
9.13 &
9.20 &
9.44 &
9.80 &
9.09 &
9.12 &
9.28 &
9.76 &
9.07 &
9.07 &
9.28 &
9.67 &
9.12 &
8.97 &
9.16 &
9.50\tabularnewline
\hline 
\end{tabular}
\end{table*}

We now proceed to analyse the performances of the filters for different
clutter parameters. The RMS-GOSPA errors considering all time steps
in the simulation for several values of $a^{C}$ and $\overline{\lambda}^{C}$
($p^{D}=0.9$) are shown in Table \ref{tab:RMS-GOSPA-errors-across_clutter}.
The A-PMB filter is most of the times the best performing filter closely
followed by the A-PMBM filter. The standard PMB and PMBM filters tend
to work better in comparison with A-PMBM and A-PMB filters for low
value of $a^{C}$. This is expected as, for values of $a^{C}$ close
to 1, the negative binomial and Poisson distributions become alike,
see Figure \ref{fig:KLD-NB-Poisson}, and it becomes beneficial to
use the standard filters with PPP clutter due to the improved hypothesis
representation, see Section \ref{subsec:Number-of-global-hypotheses}.
In fact, for $a^{C}=2$, the PMBM filter is the best performing filter
overall. The $\delta$-GLMB filter has higher errors than the previous
filters, and is followed by the MBM filter. PHD filters are the filters
with the worst performance. 

\subsection{Extended targets with independent clutter sources\label{subsec:Extended-targets-independent_clutter}}

In this section, we consider a scenario with extended targets where
the clutter is the union of independent PPP and a number of stationary
extended sources, see Appendix \ref{sec:AppendixC}. We compare the
following filters with their gamma Gaussian inverse-Wishart (GGIW)
implementations: the standard extended target PMBM and PMB filter
(with PPP clutter assumption) \cite{Granstrom20,Xia22}, the extended
target PMBM and PMB filters with arbitrary clutter, also referred
to as A-PMBM and A-PMB. We also compare with the GGIW implementations
of the MBM filter \cite{Xia22} and the $\delta$-GLMB filter \cite{Beard16}. 

All PMBM-PMB filters have been implemented using a two-step clustering
and assignment approach to select relevant global hypotheses in the
update of each previous global hypotheses \cite{Granstrom17}. Specifically,
we first apply the density-based spatial clustering of applications
with noise (DB-SCAN) \cite{Ester96} using 25 different distance values
equally spaced between 0.1 and 5 to obtain a set of different measurement
partitions, and then for each measurement partition and global hypothesis
$a$, we apply Murty's algorithm \cite{Murty68} to find the $\lceil20w_{k|k}^{a}\rceil$
best cluster-to-Bernoulli assignments. The rest of the parameters
are: maximum number of global hypotheses $N_{h}=20$, threshold for
MBM pruning $10^{-2}$, threshold for pruning the PPP weights $\Gamma_{p}=10^{-3}$,
threshold for pruning Bernoulli densities $\Gamma_{b}=10^{-3}$, estimator
1 with threshold 0.4, and ellipsoidal gating with threshold 20. The
$\delta$-GLMB filter has been implemented as in \cite{Beard16} with
a maximum number of global hypotheses $N_{h}=20$. 

The target state in GGIW implementation is $x=(\gamma,\xi,X)$, where
$\gamma$ represents the expected number of measurements per target,
$\xi=[p_{x},v_{x},p_{y},v_{y}]^{\top}$ contains the target current
position and velocity, and $X$ is a positive definite matrix with
size 2, describing the target ellipsoidal shape. The dynamical model
for the kinematic state $\xi$ is the same as the one used for point
target tracking, and states $\gamma$ and $X$ remain unchanged over
time. The probability of survival $p^{S}=0.99$. 

\begin{figure}[!t]
\centering \includegraphics[width=1\linewidth]{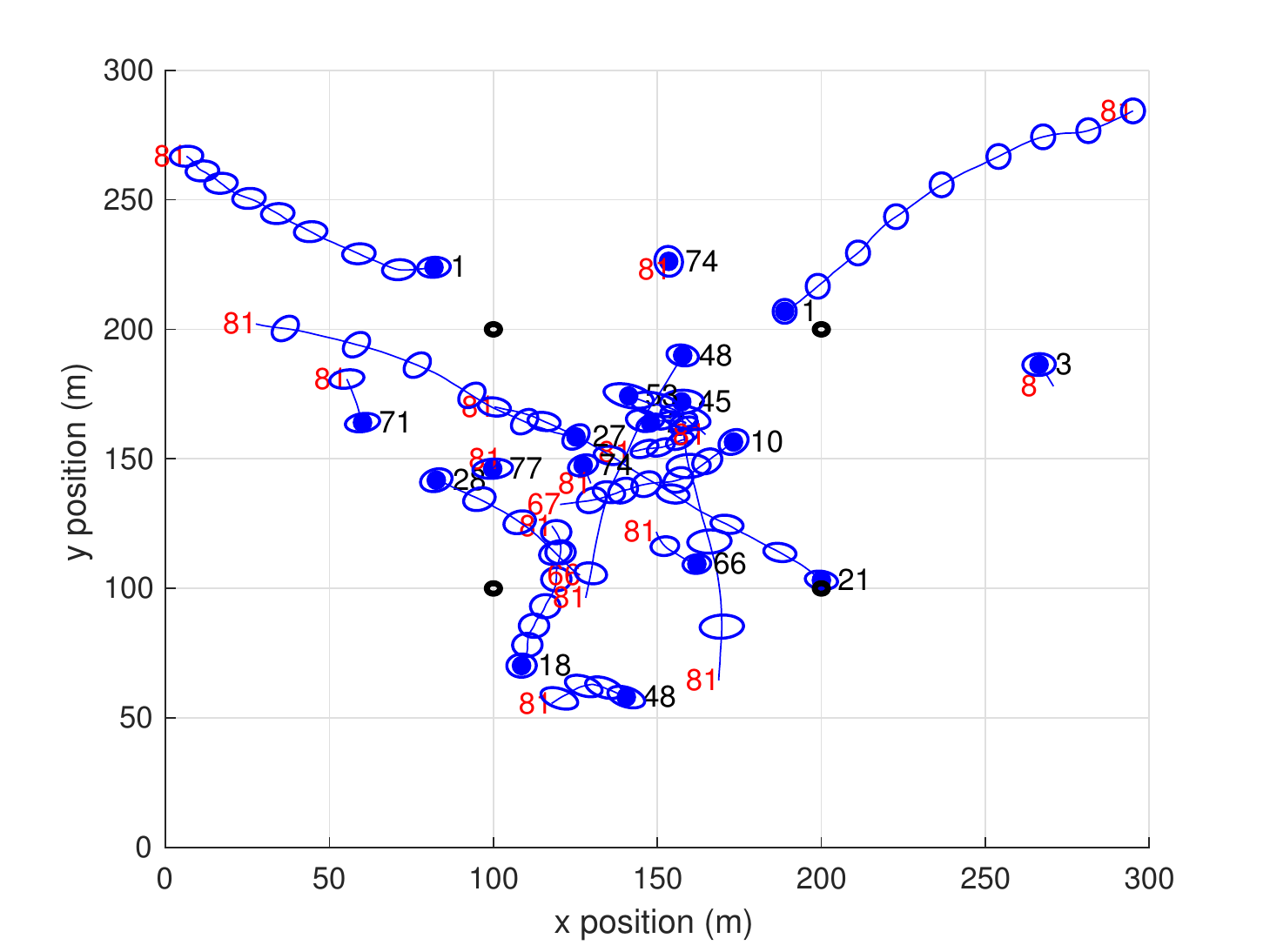}
\caption{Ground truth set of trajectories for the extended target scenario.
Initial trajectory positions are marked with filled circles and their
extents (in blue ellipses) are shown every ten times. The four stationary
sources are marked with black circles. The black numbers next to birth
positions indicate the time of birth and the red numbers the last
time a target is alive. At time step 21, a target is born adjacent
to the stationary source at $[200,100]^{\top}\,(\mathrm{m})$. At
time step 48, there is a target born nearby an existing target.}
\label{fig_ground_truth_eot}
\end{figure}

The birth process is a PPP with a GGIW intensity with weight $w_{k}^{b}=0.1$
for all time steps. Its GGIW density consists of a gamma distribution
with mean 5 and shape 100, a Gaussian distribution with mean $\bar{x}_{k}^{b}=[150\,(\mathrm{m)},0\,(\mathrm{m/s)},150\,(\mathrm{m)},0\,(\mathrm{m/s)}]^{\top}$
and covariance matrix $P_{k}^{b}=\text{diag}([50^{2}\,(\mathrm{m}^{2}),1\,(\mathrm{m}^{2}/\mathrm{s}^{2}),50^{2}\,(\mathrm{m}^{2}),1\,(\mathrm{m}^{2}/\mathrm{s}^{2})])$,
and an inverse-Wishart distribution with mean $\text{diag}([4,4])$
$(\mathrm{m}^{2})$ and degrees of freedom 100. The ground truth set
of trajectories, shown in Figure \ref{fig_ground_truth_eot}, is generated
by sampling the above dynamic model for 81 time steps. The MBM and
$\delta$-GLMB use a Bernoulli birth with probability of existence
0.1 and the same mean and covariance matrix so that their PHDs match.

At each time step, a target is detected with probability $p^{D}=0.9$,
and, if it is detected, the target measurements are a PPP with Poisson
rate $5$ and single measurement density $\mathcal{N}(z;H\xi,sX+R)$
where scaling factor $s=1/4$ and measurement noise covariance $R=\text{diag}([1/4,1/4])$
$\,(\mathrm{m}^{2})$. There are four independent stationary clutter
sources, located at $[100,100]^{\top}$, $[100,200]^{\top}$, $[200,100]^{\top}$
and $[200,200]^{\top}$ $\,(\mathrm{m)}$, respectively. Each stationary
clutter source is detected with probability $0.98$, and, if it is
detected, its measurements are a PPP with Poisson rate $10$ and Gaussian
measurement density with covariance $\text{diag}([2,2])\,(\mathrm{m}^{2})$.
The additional PPP clutter is uniformly distributed in the region
of interest $A=[0,300]\times[0,300]\,(\mathrm{m}\times\mathrm{m})$
with Poisson clutter rate $\bar{\lambda}^{C}=20$.

\begin{figure}[!t]
\centering \includegraphics[width=1\linewidth]{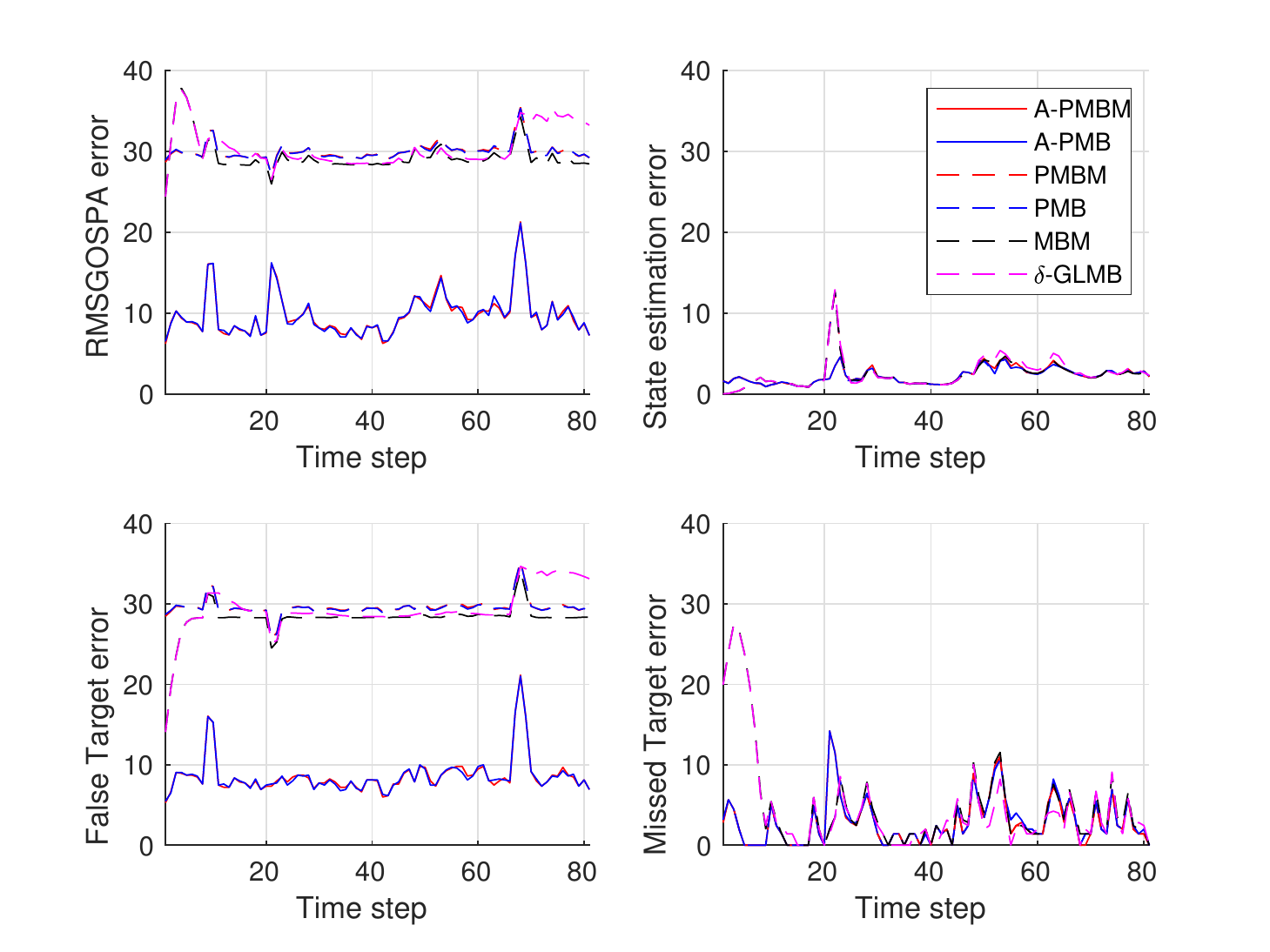}
\caption{RMS GOSPA errors and their decomposition against time for the extended
target scenario. A-PMBM and A-PMB perform considerably better than
PMBM and PMB filters.}
\label{fig_results_eot}
\end{figure}

We evaluate the performances of the filters using Monte Carlo simulation
with 100 runs. We obtain the root mean square GOSPA metric error ($p=2$,
$c=10\,\mathrm{m}$, $\alpha=2$) at each time step, where the base
measure is given by the Gaussian Wasserstein distance \cite{Yang16}.
The resulting GOSPA errors as well as their decompositions are shown
in Figure \ref{fig_results_eot}. From the results, we can see that
the standard PMBM, PMB, $\delta$-GLMB and MBM filters present high
false target errors as they treat stationary clutter sources as targets.
As a comparison, both A-PMBM and A-PMB manage to distinguish stationary
clutter sources from moving objects. The estimation performances of
A-PMBM and A-PMB are rather similar, and they report the birth of
the newborn target at time step 21, which is adjacent to one of the
stationary clutter sources, with delay. 

The computational times in seconds to run one Monte Carlo simulations
on an Intel(R) Xeon(R) Gold 6244 CPU are 80.5 (A-PMBM), 24.6 (A-PMB),
107.2 (PMBM), 25.5 (PMB), 91.5 (MBM) and 414.5 ($\delta$-GLMB). A-PMBM
and A-PMB are faster than their versions without arbitrary clutter,
as these initiate more Bernoulli components. A-PMB is faster than
A-PMBM as it does not propagate the full multi-Bernoulli mixture through
filtering. The $\delta$-GLMB filter is the slowest filter due to
its global hypothesis structure and additional calculations.

\section{Conclusions\label{sec:Conclusions}}

This paper has proved that, regardless of the distribution of the
target-generated measurements and the clutter, the posterior is a
PMBM for the standard multi-target dynamic models with PPP birth.
This result implies that, if the birth model is instead multi-Bernoulli,
the posterior is MBM, which can be labelled, and also written in MBM$_{01}$/$\delta$-GLMB
form. We have also developed an implementation of the PMBM filter,
and the corresponding PMB filter, for point targets and arbitrary
clutter based on Gibbs sampling to obtain meaningful data association
hypotheses. 

We have evaluated the filter in two scenarios. First, we present a
point-target scenario in which clutter is an IID cluster process with
negative binomial cardinality distribution, we show the performance
benefits of the A-PMBM and A-PMB filters compared with filters with
PPP clutter. Second, we present the performance benefits of the A-PMBM
and A-PMB filters in an extended-target scenario in which clutter
is the union of a PPP process plus a number of independent sources
of clutter. 

Given the generality of the measurement model, there are many lines
of future work, for example, developing specific models for target-generated
measurements and clutter tailored to different applications, estimating
their parameters, and implementing the corresponding PMBM/PMB/MBM/MB
filters. It is also direct to extend these results to the multi-sensor
case. Another line of future work is the implementation of these filters
for a large number of targets, which usually requires the use of clustering
and efficient data structures \cite{Reid79,Fontana_arxiv22,Beard20,Roy97}.

\bibliographystyle{IEEEtran}
\bibliography{8C__Trabajo_laptop_Mis_articulos_Finished_PMBM_arbitrary_clutter_Accepted_Referencias}

\cleardoublepage{}

{\LARGE{}Supplemental material: ``Poisson multi-Bernoulli mixture
filter with general target-generated measurements and arbitrary clutter''}{\LARGE\par}

\appendices{}

\section{\label{sec:AppendixA}}

This appendix proves the general PMBM filter update in Theorem \ref{thm:PMBM_update_arbitrary_clutter}
using probability generating functionals (PGFLs) \cite{Mahler_book14}.
In Section \ref{subsec:PGFLs-of-prior-measurements}, we first write
the PGFLs of the prior and the measurements given the multi-target
state. The joint PGFL of the prior and the measurements is derived
in Section \ref{subsec:Joint-PGFL}. Finally, the PGFL of the posterior
and the resulting posterior are derived in Section \ref{subsec:Updated-PGFL}. 

\subsection{PGFLs of prior and measurements\label{subsec:PGFLs-of-prior-measurements}}

Let $h\left(\cdot\right)$ be a unitless, real-valued function in
the single-target space, and $h^{X}=\prod_{x\in X}h(x)$ with $h^{\emptyset}=1$.
Then, the PGFL of the PMBM in (\ref{eq:PMBM}) is \cite{Williams15b}
\begin{align}
G_{k|k'}[h] & =\int h^{X}f_{k|k'}\left(X\right)\delta X,\\
G_{k|k'}[h] & =G_{k|k'}^{\mathrm{p}}[h]\cdot G_{k|k'}^{\mathrm{mbm}}[h],\label{eq:Hybrid_PGFL}\\
G_{k|k'}^{\mathrm{p}}[h] & =\exp\left(\langle\lambda_{k|k'},h-1\rangle\right)\propto\exp\left(\langle\lambda_{k|k'},h\rangle\right),\label{eq:PPP_PGFL}\\
G_{k|k'}^{\mathrm{mbm}}[h] & =\sum_{a\in\mathcal{A}_{k'|k}}w_{k|k'}^{a}\prod_{i=1}^{n_{k'|k}}G_{k|k'}^{i,a^{i}}[h]\label{eq:MBM_PGFL}\\
 & \propto\sum_{a\in\mathcal{A}_{k'|k}}w_{k|k'}^{0,a^{0}}\prod_{i=1}^{n_{k|k'}}\left[w_{k|k'}^{i,a^{i}}G_{k|k'}^{i,a^{i}}[h]\right]
\end{align}
where
\begin{align}
G_{k|k'}^{i,a^{i}}[h] & =1-r_{k|k'}^{i,a^{i}}+r_{k|k'}^{i,a^{i}}\langle p_{k|k'}^{i,a^{i}},h\rangle.
\end{align}

Given the multi-target state $X$, measurements from each target are
independent, and there is also independent clutter. Therefore, the
PGFL $M_{Z}[g|X]$ of the measurements given $X$ is the product of
PGFLs
\begin{align}
M_{Z}[g|X] & =M_{C}[g]\prod_{x\in X}M_{T}[g|x]
\end{align}
where $M_{T}[g|x]$ is the PGFL of $f\left(\cdot|x\right)$, $M_{C}[g]$
is the PGFL of $c\left(\cdot\right)$, and $g\left(\cdot\right)$
is a unitless, real-valued function in the single-measurement space.

\subsection{Joint PGFL\label{subsec:Joint-PGFL}}

The joint PGFL of the joint density of the set of measurements and
set of targets is \cite{Williams15b,Mahler_book14}
\begin{align}
F[g,h] & =\int\int g^{Z_{k}}h^{X_{k}}f\left(Z_{k}|X_{k}\right)f_{k|k-1}\left(X_{k}\right)\delta Z_{k}\delta X_{k}\nonumber \\
 & =\int M_{Z}[g|X_{k}]h^{X_{k}}f_{k|k-1}\left(X_{k}\right)\delta X_{k}\\
 & =M_{C}[g]G_{k|k-1}[hM_{T}[g|\cdot]]\\
 & \propto\exp\left(\langle\lambda_{k|k-1},hM_{T}[g|\cdot]\rangle\right)M_{C}[g]\sum_{a\in\mathcal{A}_{k|k-1}}w_{k|k-1}^{0,a^{0}}\nonumber \\
 & \times\prod_{i=1}^{n_{k|k-1}}\left[w_{k|k-1}^{i,a^{i}}G_{k|k-1}^{i,a^{i}}\big[hM_{T}[g|\cdot]\big]\right].\label{eq:JointPGFLFull}
\end{align}

We denote the first factor in (\ref{eq:JointPGFLFull}) as
\begin{align}
F^{0}[g,h] & =\exp\left(\langle\lambda_{k|k-1},hM_{T}[g|\cdot]\rangle\right)\label{eq:JointPGFLPoissonComponent}
\end{align}
which represents the joint PGFL of measurements (not including clutter)
and targets in the PPP, up to a proportionality constant. We also
denote
\begin{align}
F^{i,a^{i}}[g,h] & =G_{k|k-1}^{i,a^{i}}\big[hM_{T}[g|\cdot]\big]\\
 & =1-r_{k|k-1}^{i,a^{i}}+r_{k|k-1}^{i,a^{i}}\bigg\langle p_{k|k-1}^{i,a^{i}},hM_{T}[g|\cdot]\bigg\rangle,\label{eq:JointPGFLBernoulliComponent}
\end{align}
which is the joint PGFL of measurements (not including clutter) and
the $i$-th target. Then, we can write (\ref{eq:JointPGFLFull}) as
\begin{align}
F[g,h] & \propto F^{0}[g,h]M_{C}[g]\nonumber \\
 & \times\sum_{a\in\mathcal{A}_{k|k-1}}w_{k|k-1}^{0,a^{0}}\prod_{i=1}^{n_{k|k-1}}\left[w_{k|k-1}^{i,a}F^{i,a^{i}}[g,h]\right].\label{eq:JointPGFLSimplified}
\end{align}

\subsection{Updated PGFL\label{subsec:Updated-PGFL}}

We proceed to calculate the PGFL $G_{k|k}[h]$ of the updated density
$f_{k|k}\left(\cdot\right)$. This PGFL can be calculated by the set
derivative of $F[g,h]$ w.r.t. $Z_{k}$ evaluated at $g=0$ \cite[Sec. 5.8]{Mahler_book14}\cite[Eq. (25)]{Williams15b}
\begin{align}
G_{k|k}[h] & \propto\frac{\delta}{\delta Z_{k}}F[g,h]\bigg|_{g=0}.\label{eq:Update_PGLF_rule}
\end{align}

Substituting (\ref{eq:JointPGFLSimplified}) into (\ref{eq:Update_PGLF_rule})
and applying the product rule for sets derivatives \cite[Eq. (3.68)]{Mahler_book14},
we obtain
\begin{align}
 & G_{k|k}[h]\nonumber \\
 & \propto\sum_{W^{C}\uplus W_{0}\uplus\cdots\uplus W_{n_{k|k-1}}=Z_{k}}\frac{\delta}{\delta W^{C}}M_{C}[g]\frac{\delta}{\delta W_{0}}F^{0}[g,h]\nonumber \\
 & \times\sum_{a\in\mathcal{A}_{k|k-1}}w_{k|k-1}^{0,a^{0}}\prod_{i=1}^{n_{k|k-1}}\frac{\delta}{\delta W_{i}}\left(w_{k|k-1}^{i,a^{i}}F^{i,a^{i}}[g,h]\right)\bigg|_{g=0}.\label{eq:UpdatedJointPGFL}
\end{align}
In the first sum in (\ref{eq:UpdatedJointPGFL}), we decompose the
measurement set $Z_{k}$ into $n_{k|k-1}+2$ subsets. The subset $W^{C}$
represents measurements that are considered clutter, the subset $W_{0}$
represents measurements that are the first detection of an undetected
target (modelled in the PPP), and subset $W_{i}$, $i>0$, represents
measurements assigned to the $i$-th predicted Bernoulli component.
We proceed to calculate the required set derivatives. 

\subsubsection{Set derivative w.r.t. $W^{C}$ (clutter weight)}

As $M_{C}[g]$ is the PGFL of $c\left(\cdot\right)$, we can directly
see by the properties of the set derivative of a PGFL \cite[Sec. 3.5.1]{Mahler_book14}
that
\begin{align}
\frac{\delta}{\delta W^{C}}M_{C}[g]\bigg|_{g=0} & =c\left(W^{C}\right).\label{eq:set_derivative_clutter}
\end{align}

\subsubsection{Set derivative w.r.t. $W_{0}$ (PPP update)}

We calculate
\begin{align*}
\frac{\delta}{\delta W_{0}}F^{0}[g,h]\bigg|_{g=0}.
\end{align*}
This corresponds to the PGFL in \cite[Eq. (84)]{Angel21} setting
the clutter intensity equal to zero. Therefore, following \cite[Lem. 4]{Angel21},
we obtain

\begin{align}
\frac{\delta}{\delta W_{0}}F^{0}[g,h] & =F^{0}[g,h]\sum_{P\angle W_{0}}\prod_{V\in P}d_{V}[g,h]\label{eq:derivative_F_W0}
\end{align}
where 
\begin{equation}
d_{V}[g,h]=\frac{\delta}{\delta V}\left(\big\langle\lambda_{k|k-1},hM_{T}[g|\cdot]\big\rangle\right)\label{eq:PHDUpdatePGFLDerivdVdefn}
\end{equation}
and $\sum_{P\angle W_{0}}$ denotes the sum over all partitions $P$
of $W_{0}$. 

We evaluate the first factor in (\ref{eq:derivative_F_W0}), $F^{0}[g,h]$,
at $g=0$ resulting in 
\begin{align}
F^{0}[0,h] & =\exp\left(\big\langle\lambda_{k|k-1},hf\left(\emptyset|\cdot\right)\big\rangle\right)
\end{align}
where we have used that $M_{T}[0|\cdot]=f\left(\emptyset|\cdot\right)$.
This is proportional to the PGFL of a PPP with intensity (\ref{eq:updated_PPP}).
Then, the following lemma indicates the distributions resulting from
evaluating (\ref{eq:PHDUpdatePGFLDerivdVdefn}) at $g=0$. The lemma
is obtained by setting the clutter intensity to zero in \cite[Lem. 5]{Angel21}.
\begin{lem}
\label{lem:PPPUpdateComponents} The update of the PGFL of the PPP
prior with measurement subset $V$, $|V|>0$, 
\[
w_{k|k}^{V}G_{k|k}^{V}[h]=d_{V}[g,h]\Big|_{g=0}
\]
are PGFLs of weighted Bernoulli distributions with the form \cite[Lem. 2]{Williams15b}
\begin{equation}
f_{k|k}^{V}(X)=w_{k|k}^{V}\times\begin{cases}
1-r_{k|k}^{V} & X=\emptyset\\
r_{k|k}^{V}p_{k|k}^{V}(x) & X=\{x\}\\
0 & \left|X\right|>1
\end{cases}
\end{equation}
where
\begin{align}
w_{k|k}^{V} & =l_{k|k}^{V},\label{eq:PoissonClusterUpdateWeight}\\
l_{k|k}^{V} & =\bigg\langle\lambda_{k|k-1},f\left(V|\cdot\right)\bigg\rangle,\\
r_{k|k}^{V} & =\frac{l_{k|k}^{V}}{w_{k|k}^{V}}=1,\label{eq:PoissonClusterUpdatePExist}\\
p_{k|k}^{V}(x) & =\frac{f\left(V|x\right)\lambda_{k|k-1}\left(x\right)}{l_{k|k}^{V}}.\quad\square\label{eq:PoissonClusterUpdatePDF}
\end{align}
\end{lem}
This lemma indicates how to create the new Bernoulli components in
Theorem \ref{thm:PMBM_update_arbitrary_clutter}.

\subsubsection{Set derivative w.r.t. $W_{i}$ (Bernoulli update)}

The third factor is analogous to the one in \cite[App. A.C.1]{Angel21},
which yields
\begin{align}
\frac{\delta}{\delta W_{i}}F^{i,a^{i}}[g,h]\bigg|_{g=0} & =\begin{cases}
F^{i,a^{i}}[g,h] & W_{i}=\emptyset\\
r_{k|k-1}^{i,a^{i}}\Big\langle f_{k|k-1}^{i,a^{i}},hf\left(W_{i}|\cdot\right)\Big\rangle & W_{i}\neq\emptyset.
\end{cases}\label{eq:Derivative_Wi}
\end{align}

As shown in \cite[Lem. 2]{Williams15b}, this equation represents
weighted Bernoulli densities with the parameters indicated in Theorem
\ref{thm:PMBM_update_arbitrary_clutter} (update of previous Bernoulli),
see also \cite[Lem. 3]{Angel21}.

\subsubsection{Final expression}

Substituting (\ref{eq:set_derivative_clutter}), (\ref{eq:derivative_F_W0})
and (\ref{eq:Derivative_Wi}) into (\ref{eq:UpdatedJointPGFL}), we
obtain
\begin{align}
 & G_{k|k}[h]\nonumber \\
 & \propto F^{0}[0,h]\sum_{W^{C}\uplus W_{0}\uplus\cdots\uplus W_{n_{k|k-1}}=Z_{k}}c\left(W^{C}\right)\nonumber \\
 & \times\sum_{P\angle W_{0}}\prod_{V\in P}w_{k|k}^{V}G_{k|k}^{V}[h]\nonumber \\
 & \times\sum_{a\in\mathcal{A}_{k|k-1}}w_{k|k-1}^{0,a^{0}}\prod_{i=1}^{n_{k|k-1}}\left(w_{k|k-1}^{i,a^{i}}\frac{\delta}{\delta W_{i}}F^{i,a^{i}}[g,h]\bigg|_{g=0}\right).
\end{align}
In this expression, $F^{0}[0,h]$ represents the PPP of undetected
targets with intensity (\ref{eq:updated_PPP}). Then, for each predicted
global hypothesis, we generate a new global hypotheses by partitioning
the measurements into $W^{C},W_{0},\cdots,W_{n_{k|k-1}}$ and going
through the partitions of $W_{0}.$ The weight of the clutter hypothesis
is given by $c\left(W^{C}\right)w_{k|k-1}^{0,a^{0}}$, which results
in (\ref{eq:clutter_no_measurement_update}) and (\ref{eq:clutter_measurement_update}).
Then, for each new global hypothesis, we have a multi-Bernoulli, where
the previous Bernoulli densities are updated as in (\ref{eq:Miss_measurement})-(\ref{eq:update_density}),
and the new Bernoulli components as in (\ref{eq:new_Bernoulli_likelihood})-(\ref{eq:new_Bernoulli_density}).
We then finish the proof of Theorem \ref{thm:PMBM_update_arbitrary_clutter},
by writing the updated density in a track-oriented form.

\section{\label{sec:AppendixC}}

In this appendix, we write the expression of the PMBM update when
clutter is the union of Poisson clutter and $n_{c}$ independent sources
of clutter. That is, the clutter density is 
\begin{align}
c\left(Z_{c}\right) & =\sum_{\uplus_{l=0}^{n_{c}}Z^{l}=Z_{c}}c^{0}\left(Z^{0}\right)\prod_{i_{c}=1}^{n_{c}}c^{i_{c}}\left(Z^{i_{c}}\right)\label{eq:Poisson_n_independent_clutter}
\end{align}
where
\begin{align*}
c^{0}\left(Z^{0}\right) & =e^{-\int\lambda^{C}\left(z\right)dz}\prod_{z\in Z^{0}}\lambda^{C}\left(z\right)
\end{align*}
where $c^{0}\left(\cdot\right)$ is the density of the PPP clutter
with intensity $\lambda^{C}\left(\cdot\right)$, and $c^{i_{c}}\left(\cdot\right)$
is the density of the $i_{c}$ source of clutter.

We can directly use Theorem \ref{thm:PMBM_update_arbitrary_clutter}
to obtain the updated PMBM density. With this approach, we consider
a single local hypothesis with clutter measurements. If the clutter
density has form (\ref{eq:Poisson_n_independent_clutter}), we can
alternatively define a local hypothesis $a^{i}\in\left\{ 1,...,h_{k|k'}^{i}\right\} $
for each Bernoulli component $i>0$, and also a local hypothesis $a^{0,i_{c}}\in\left\{ 1,...,h_{k|k'}^{0,i_{c}}\right\} $
for the $i$-th clutter source. A global hypothesis is $a=\left(a^{0,1},...,a^{0,n_{c}},a^{1},...,a^{n_{k|k'}}\right)\in\mathcal{A}_{k|k'}$,
where $\mathcal{A}_{k|k'}$ is the set of global hypotheses. That
is, as in PMBM filters with PPP clutter, the \cite{Williams15b,Granstrom20,Angel21},
the creation of new Bernoulli components will take into account the
PPP clutter, and then, we also have additional sources of clutter.

The weight of a global hypothesis $a$ is
\begin{equation}
w_{k|k'}^{a}\propto\prod_{i_{c}=1}^{n_{c}}w_{k|k'}^{0,i_{c},a^{0,i_{c}}}\prod_{i=1}^{n_{k|k'}}w_{k|k'}^{i,a^{i}}.\label{eq:global_weights-P_n_independent}
\end{equation}

At time step zero, the filter is initiated with $n_{0|0}=0$, $w_{0|0}^{0,i_{c}}=1$,
$h_{0|0}^{0,i_{c}}=1$, $\mathcal{M}_{0}^{0,i_{c}}=\emptyset$. The
update is provided in the following theorem.
\begin{thm}
\label{thm:PMBM_update-Poisson_n_independent_clutter}Assume the predicted
density $f_{k|k-1}\left(\cdot\right)$ is a PMBM of the form (\ref{eq:PMBM}).
Then, the updated density $f_{k|k}\left(\cdot\right)$ with set $Z_{k}=\left\{ z_{k}^{1},...,z_{k}^{m_{k}}\right\} $
is a PMBM with the following parameters. The number of Bernoulli components
is $n_{k|k}=n_{k|k-1}+2^{m_{k}}-1$. The intensity of the PPP is
\begin{align}
\lambda_{k|k}\left(x\right) & =f\left(\emptyset|x\right)\lambda_{k|k-1}\left(x\right).\label{eq:updated_PPP-1}
\end{align}

Let $Z_{k}^{1},...,Z_{k}^{2^{m_{k}}-1}$ be the non-empty subsets
of $Z_{k}$. The number of updated local clutter hypotheses for the
$i_{c}$-th clutter source is $h_{k|k}^{0,i_{c}}=2^{m_{k}}h_{k|k-1}^{0,i_{c}}$
such that a new local clutter hypothesis is included for each previous
local clutter hypothesis and either a misdetection or an update with
a non-empty subset of $Z_{k}$. The updated local hypothesis with
0 measurements for the $i$-th clutter source, $a^{0,i_{c}}\in\left\{ 1,...,h_{k|k-1}^{0,i_{c}}\right\} $,
has parameters
\begin{align*}
\mathcal{M}_{k}^{0,i_{c},a^{0,i_{c}}} & =\mathcal{M}_{k-1}^{0,i_{c},a^{0}},\\
w_{k|k}^{0,i_{c},a^{0,i_{c}}} & =w_{k|k-1}^{0,i_{c},a^{0,i_{c}}}c^{i_{c}}\left(\emptyset\right).
\end{align*}
For a previous local clutter hypothesis $\widetilde{a}^{0,i_{c}}\in\left\{ 1,...,h_{k|k-1}^{0,i_{c}}\right\} $
for the $i_{c}$ clutter source in the predicted density, the new
local clutter hypothesis generated by a a set $Z_{k}^{j}$ has $a^{0,i_{c}}=\widetilde{a}^{0,i_{c}}+h_{k|k-1}^{0,i_{c}}j$,
\begin{align*}
\mathcal{M}_{k}^{0,i_{c},a^{0,i_{c}}} & =\mathcal{M}_{k-1}^{0,i_{c},\widetilde{a}^{0,i_{c}}}\cup\left\{ \left(k,p\right):z_{k}^{p}\in Z_{k}^{j}\right\} ,\\
w_{k|k}^{0,i_{c},a^{0,i_{c}}} & =w_{k|k-1}^{0,i_{c},a^{0,i_{c}}}c^{i_{c}}\left(Z_{k}^{j}\right).
\end{align*}
For Bernoulli components continuing from previous time steps $i\in\left\{ 1,...,n_{k|k-1}\right\} $,
a new local hypothesis is included for each previous local hypothesis
and either a misdetection or an update with a non-empty subset of
$Z_{k}$. The updated number of local hypotheses is $h_{k|k}^{i}=2^{m_{k}}h_{k|k-1}^{i}$.
For missed detection hypotheses, $i\in\left\{ 1,...,n_{k|k-1}\right\} $,
$a^{i}\in\left\{ 1,...,h_{k|k-1}^{i}\right\} $, we obtain
\begin{align}
\mathcal{M}_{k}^{i,a^{i}} & =\mathcal{M}_{k-1}^{i,a^{i}},\label{eq:Miss_measurement-1}\\
l_{k|k}^{i,a^{i},\emptyset} & =\big\langle f_{k|k-1}^{i,a^{i}},f\left(\emptyset|\cdot\right)\big\rangle,\label{eq:Miss_likelihood-1}\\
w_{k|k}^{i,a^{i}} & =w_{k|k-1}^{i,a^{i}}\left[1-r_{k|k-1}^{i,a^{i}}+r_{k|k-1}^{i,a^{i}}l_{k|k}^{i,a^{i},\emptyset}\right],\label{eq:Miss_weight-1}\\
r_{k|k}^{i,a^{i}} & =\frac{r_{k|k-1}^{i,a^{i}}l_{k|k}^{i,a^{i},\emptyset}}{1-r_{k|k-1}^{i,a^{i}}+r_{k|k-1}^{i,a^{i}}l_{k|k}^{i,a^{i},\emptyset}},\label{eq:Miss_existence-1}\\
f_{k|k}^{i,a^{i}}(x) & =\frac{f\left(\emptyset|x\right)f_{k|k-1}^{i,a^{i}}(x)}{l_{k|k}^{i,a^{i},\emptyset}}.\label{eq:Miss_density-1}
\end{align}

Let $Z_{k}^{1},...,Z_{k}^{2^{m_{k}}-1}$ be the non-empty subsets
of $Z_{k}$. For a Bernoulli component $i\in\left\{ 1,...,n_{k|k-1}\right\} $
with a local hypothesis $\widetilde{a}^{i}\in\left\{ 1,...,h_{k|k-1}^{i}\right\} $
in the predicted density, the new local hypothesis generated by a
set $Z_{k}^{j}$ has $a^{i}=\widetilde{a}^{i}+h_{k|k-1}^{i}j$, $r_{k|k}^{i,a^{i}}=1$,
and
\begin{align}
\mathcal{M}_{k}^{i,a^{i}} & =\mathcal{M}_{k-1}^{i,\widetilde{a}^{i}}\cup\left\{ \left(k,p\right):z_{k}^{p}\in Z_{k}^{j}\right\} ,\\
l_{k|k}^{i,a^{i},Z_{k}^{j}} & =\bigg\langle f_{k|k-1}^{i,\widetilde{a}^{i}},f\left(Z_{k}^{j}|\cdot\right)\bigg\rangle,\\
w_{k|k}^{i,a^{i}} & =w_{k|k-1}^{i,\widetilde{a}^{i}}r_{k|k-1}^{i,\widetilde{a}^{i}}l_{k|k}^{i,a^{i},Z_{k}^{j}},\label{eq:update_weight-1}\\
f_{k|k}^{i,a^{i}}(x) & =\frac{f\left(Z_{k}^{j}|x\right)f_{k|k-1}^{i,\widetilde{a}^{i}}(x)}{l_{k|k}^{i,a^{i},Z_{k}^{j}}}.\label{eq:update_density-1}
\end{align}
For the new Bernoulli component initiated by subset $Z_{k}^{j}$,
whose index is $i=n_{k|k-1}+j$, we have two local hypotheses ($h_{k|k}^{i}=2$),
one corresponding to a non-existent Bernoulli density
\begin{equation}
\mathcal{M}_{k}^{i,1}=\emptyset,\;w_{k|k}^{i,1}=1,\;r_{k|k}^{i,1}=0,
\end{equation}
and the other with
\begin{align}
\mathcal{M}_{k}^{i,2} & =\left\{ \left(k,p\right):z_{k}^{p}\in Z_{k}^{j}\right\} ,\\
l_{k|k}^{Z_{k}^{j}} & =\bigg\langle\lambda_{k|k-1},f\left(Z_{k}^{j}|\cdot\right)\bigg\rangle,\\
w_{k|k}^{i,2} & =\delta_{1}\left[|Z_{k}^{j}|\right]\left[\prod_{z\in Z_{k}^{j}}\lambda^{C}\left(z\right)\right]+l_{k|k}^{Z_{k}^{j}},\label{eq:new_Bernoulli_weight-1}\\
r_{k|k}^{i,2} & =\frac{l_{k|k}^{Z_{k}^{j}}}{w_{k|k}^{i,a^{i}}},\label{eq:new_Bernoulli_existence-1}\\
f_{k|k}^{i,2}(x) & =\frac{f\left(Z_{k}^{j}|x\right)\lambda_{k|k-1}(x)}{l_{k|k}^{Z_{k}^{j}}}.\quad\square\label{eq:new_Bernoulli_density-1}
\end{align}
\end{thm}
The proof is equivalent to the one of Theorem \ref{thm:PMBM_update_arbitrary_clutter}
in Appendix \ref{sec:AppendixA}, but adding $n_{c}$ sources of clutter
and a PPP clutter source. The PPP clutter source is incorporated in
the update for new Bernoulli components as in \cite{Angel21}. The
main differences with Theorem \ref{thm:PMBM_update_arbitrary_clutter}
are that now there are updates with detection and misdetection hypotheses
with the $n_{c}$ clutter sources, and that the intensity of the clutter
affects (\ref{eq:new_Bernoulli_weight-1}). This implies that the
probability of existence for new Bernoulli components can be lower
than 1. It should be noted that if in Theorem \ref{thm:PMBM_update-Poisson_n_independent_clutter},
we set $n_{c}=1$ and $\lambda^{C}\left(\cdot\right)=0$, we recover
Theorem \ref{thm:PMBM_update_arbitrary_clutter}.
\end{document}